%% file: main.tex
\def\draft{0}

\documentclass[11pt]{article}
\usepackage{macros}

\newif\ifsubmission
\submissiontrue

\title{Continuous LWE is as Hard as LWE\\ {\Large \& Applications to Learning Gaussian Mixtures}}
\author{Aparna Gupte\thanks{Research supported by the Keel Foundation Undergraduate Research and Innovation Scholarship.}\\MIT\\\texttt{agupte@mit.edu} \and Neekon Vafa\thanks{Research supported by NSF fellowship DGE-1745302 and by the grants of the third author.}\\MIT\\\texttt{nvafa@mit.edu} \and Vinod Vaikuntanathan\thanks{Research supported in part by DARPA under Agreement No. HR00112020023, a grant from the MIT-IBM Watson AI, a grant from Analog Devices, a Microsoft Trustworthy AI grant, and a Thornton Family Faculty Research Innovation Fellowship from MIT. Any opinions, findings and conclusions or recommendations expressed in this material are those of the author(s) and do not necessarily reflect the views of the United States Government or DARPA.}\\MIT\\\texttt{vinodv@mit.edu}
}

\date{}

\def\LWE{\mathsf{LWE}}

\def\CLWE{\mathsf{CLWE}}
\def\hCLWE{\mathsf{hCLWE}}

\def\cD{\mathcal{D}}
\def\veca{\mathbf{a}}
\def\vecb{\mathbf{b}}
\def\vecc{\mathbf{c}}
\def\vecv{\mathbf{v}}
\def\vecw{\mathbf{w}}
\def\vecx{\mathbf{x}}
\def\vecy{\mathbf{y}}
\def\vecz{\mathbf{z}}

\def\matB{\mathbf{B}}

\def\vecs{\mathbf{s}}
\def\vecu{\mathbf{u}}
\def\vece{\mathbf{e}}

\def\vecf{\mathbf{f}}
\def\vecg{\mathbf{g}}
\def\bmu{\text{\boldmath$\mu$}}
\def\secp{\lambda}
\def\poly{\mathsf{poly}}
\def\Z{\mathbb{Z}}
\def\Zq{\Z_q}
\def\R{\mathbb{R}}

\def\S{\mathcal{S}}

\def\minentropy{H_{\infty}}

\begin{document}

\maketitle

\input{abstract.tex}

\ifsubmission
\thispagestyle{empty}
\newpage
\pagenumbering{roman}
\tableofcontents
\clearpage
\pagenumbering{arabic}

\input{intro.tex}

\input{prelims.tex}

\input{ksparse.tex}

\input{LWEtoCLWE.tex}

\input{gmm.tex}

\section*{Acknowledgment}
We thank the anonymous reviewers for their valuable feedback on previous versions of our write-up, especially one reviewer for pointing out an error in Lemma~\ref{lemma:clwe-to-hclwe}.
\bibliographystyle{alpha}
\bibliography{refs}

\appendix

\input{alternate.tex}

\input{lowsample-alg.tex}
\input{CLWEtoLWE.tex}

\end{document}

%% file: abstract.tex
\begin{abstract}
We show direct and conceptually simple reductions between the classical learning with errors (LWE) problem and its continuous analog, CLWE (Bruna, Regev, Song and Tang, STOC 2021). This allows us to bring to bear the powerful machinery of LWE-based cryptography to the applications of CLWE. For example, we obtain the hardness of CLWE under the \emph{classical} worst-case hardness of the gap shortest vector problem. Previously, this was known only under \emph{quantum} worst-case hardness of lattice problems. More broadly, with our reductions between the two problems, any future developments to LWE will also apply to CLWE and its downstream applications.

As a concrete application, we show an improved hardness result for density estimation for mixtures of Gaussians. In this computational problem, given sample access to a mixture of Gaussians, the goal is to output a function that estimates the density function of the mixture. Under the (plausible and widely believed) exponential hardness of the classical LWE problem, we show that Gaussian mixture density estimation in $\mathbb{R}^n$ with roughly $\log n$ Gaussian components given $\mathsf{poly}(n)$ samples requires time quasi-polynomial in $n$. Under the (conservative) polynomial hardness of LWE, we show hardness of density estimation for $n^{\epsilon}$ Gaussians for any constant $\epsilon > 0$, which improves on Bruna, Regev, Song and Tang (STOC 2021), who show hardness for at least $\sqrt{n}$ Gaussians under polynomial (quantum) hardness assumptions.
    
Our key technical tool is a reduction from classical LWE to LWE with $k$-sparse secrets where the multiplicative increase in the noise is only $O(\sqrt{k})$, independent of the ambient dimension $n$.
\end{abstract}

%% file: intro.tex
\section{Introduction}
\label{sec:intro}

The learning with errors (LWE) problem~\cite{regev2009lattices} is a versatile average-case problem with connections to lattices, cryptography, learning theory and game theory. Given a sequence of noisy linear equations $(\mathbf{a}, b \approx \langle \mathbf{a},\mathbf{s}\rangle \bmod{q})$ over a ring $\mathbb{Z}/q\mathbb{Z}$, the LWE problem asks to recover the secret vector $\mathbf{s}$ (and the decisional version of the problem asks to distinguish between LWE samples and uniformly random numbers mod $q$). Starting from the seminal work of Regev, who showed that a polynomial-time algorithm for LWE will give us a polynomial-time {\em quantum} algorithm for widely studied worst-case lattice problems, there has been a large body of work showing connections between LWE and lattice problems~\cite{Peikert09,brakerski2013classical}. Ever since its formulation in 2005, LWE has unlocked a wealth of applications in cryptography ranging from fully homomorphic encryption~\cite{BV14} to attribute-based encryption~\cite{GVW15} to, most recently, succinct non-interactive argument systems for all of P~\cite{CJJ21}. LWE-based cryptosystems lie at the center of efforts by the National Institute of Standards and Technology (NIST) to develop post-quantum cryptographic standards. LWE has also had applications to learning theory, in the form of hardness results for learning intersections of halfspaces~\cite{KlivansS09}, and in game theory, where the hardness of LWE implies the hardness of the complexity class PPAD~\cite{JawaleKKZ21}. Finally, LWE enjoys remarkable structural properties such as leakage-resilience~\cite{goldwasser2010robustness}.

Motivated by applications to learning problems, Bruna, Regev, Song and Tang~\cite{CLWE} recently introduced a continuous version of LWE which they called CLWE. (In the definition below and henceforth, $\mathcal{N}(\boldsymbol{\mu},\Sigma)$ is the multivariate normal distribution with mean $\boldsymbol{\mu}$ and covariance matrix $\Sigma$ where the probability of a point $\mathbf{x} \in \mathbb{R}^n$ is proportional to $e^{-\frac{1}{2} (\mathbf{x} - \boldsymbol{\mu})^T \Sigma^{-1} (\mathbf{x} - \boldsymbol{\mu})}$.)

\begin{definition}[CLWE Distribution \cite{CLWE}, rescaled]
Let $\gamma, \beta \in \R$, and let $\S$ be a distribution over unit vectors in $\R^n$. Let $\CLWE(m, \S, \gamma, \beta)$ be the distribution given by sampling $\veca_1, \cdots, \veca_m \sim \Norm(\mathbf{0}, I_{n \times n})$, $\vecw \sim \S, e_1, \cdots, e_m \sim \mathcal{N}(0, \beta^2)$ and outputting 
$$\big(\veca_i, b_i := \gamma \cdot \inner{\veca_i, \vecw} + e_i \bmod{1} \big)_{i=1}^m.$$ 
Unless otherwise specified, $\S$ is taken to be the uniform distribution over all unit vectors in $\R^n$. We refer to $n$ as the dimension and $m$ as the number of samples. 
\end{definition}

The search CLWE problem asks to find the secret vector $\mathbf{w}$ given CLWE samples, whereas the decisional CLWE problem asks to distinguish between samples from the CLWE distribution and samples with standard normal $\veca_i$ (just like the CLWE distribution) but now with independent $b_i$ that are distributed uniformly between $0$ and $1$.

Bruna et al.~\cite{CLWE} showed the hardness of the CLWE problem, assuming the worst-case {\em quantum} hardness of approximate shortest vector problems on lattices (such as \textsf{gapSVP} and \textsf{SIVP}). 
Aside from being quantum, the reduction makes non-black-box use of the rather involved techniques from \cite{regev2009lattices,PeikertRS17}. A natural question is whether CLWE has a {\em classical} reduction from worst-case lattice problems, in analogy with such reductions in the context of LWE~\cite{Peikert09,brakerski2013classical}. An even better outcome would be if we can ``piggyback'' on the rich literature on worst-case to average-case reductions for LWE, without opening the box, hopefully resulting in a conceptually simple worst-case to average-case connection for CLWE. The conceptually clean way to accomplish all of this would be to come up with a {\em direct} reduction from LWE to CLWE, a problem that was explicitly posed in the recent work of Bogdanov, Noval, Hoffman and Rosen~\cite{BNHR22}.

Our main conceptual contribution is a direct and simple reduction from LWE to CLWE. When combined with Regev~\cite{regev2009lattices}, our reduction immediately gives an alternate proof of CLWE hardness assuming worst-case quantum hardness of lattice problems, reproving one of the main results of Bruna et al. \cite{CLWE}. As another immediate application, by combining with the classical reduction from worst-case lattice problems to LWE~\cite{brakerski2013classical}, we obtain \emph{classical} worst-case hardness of CLWE. Our main reduction also allows us to unlock powerful structural results on LWE~\cite{goldwasser2010robustness,brakerski2013classical,Mic18,BD20} and derive improved hardness results for learning mixtures of Gaussians with $(\log n)^{1 + \epsilon}$ Gaussians instead of $\Omega(\sqrt{n})$ in \cite{CLWE} (for arbitrary $\epsilon > 0$). We now describe these results in turn.

\subsection{Continuous LWE is as Hard as LWE} 

Our main result is a direct and conceptually simple reduction from LWE to CLWE. Recall that in the decisional LWE problem~\cite{regev2009lattices}, we are given $m$ samples of the form $(\veca_i, b_i := \langle \veca_i, \vecs\rangle + e_i \bmod{q})$ where $\veca_i \sim (\mathbb{Z}/q\mathbb{Z})^n$ is uniformly random, $\vecs \in \Z^n$ is the LWE secret vector, and the errors $e_i \sim \mathcal{N}(0,\sigma^2)$ are chosen from the one-dimensional Gaussian with standard deviation $\sigma$. The decisional LWE assumption (parameterized by $n,m,q$ and $\sigma$) postulates that these samples are computationally indistinguishable from i.i.d. samples in $(\mathbb{Z}/q\mathbb{Z})^n \times \mathbb{R}/q\mathbb{Z}$.

\begin{theorem}[Informal Version of Theorem~\ref{thm:fixed-norm-lwe-clwe}]\label{thm:main-lwe-clwe-reduction-informal}
Let $\S = \S_r$ be an arbitrary distribution over $\Z^n$ whose support consists of vectors with $\ell_2$-norm exactly $r$. Then, for
\begin{align*}
\gamma &= \tilde{O}(r) \text{  and  } \beta = O \left(\frac{ \sigma}{q} \right),
\end{align*}
(where $\tilde{O}(\cdot)$ hides various poly-logarithmic factors), there is a dimension-preserving and sample-preserving polynomial-time reduction from decisional LWE, with parameters $n,m,q, \sigma$ and secret distribution $\S$, to decisional CLWE with parameters $n,m,\gamma$ and $\beta$, as long as $\sigma \gg r$.
\end{theorem}

Our main reduction,  in conjunction with prior work, immediately gives us a number of corollaries. First, letting $\mathcal{S}$ be the uniform distribution on $\{-1,1\}^n$,  and invoking the hardness result for LWE with binary secrets~\cite{brakerski2013classical,Mic18,BD20}, we obtain the following corollary. (The noise blowup of $\sqrt{n}$ in the corollary below comes from the aforementioned reductions from LWE to LWE with binary secrets.)

\begin{corollary}[Informal Version of Corollary~\ref{cor:lwe-to-clwe}]\label{cor1}
For 
\begin{align*}
\gamma &= \tilde{O}\left( \sqrt{n} \right) \text{  and  } \beta = O \left(\frac{ \sigma \sqrt{n}}{q} \right),
\end{align*}
there is a polynomial (in $n$) time reduction from standard decisional LWE in dimension $\ell$, with $n$ samples, modulus $q$ and noise parameter $\sigma$, to decisional CLWE in dimension $n$ with parameters $\gamma$ and $\beta$, as long as $n \gg \ell \log_2(q)$ and $\sigma \gg 1$.
\end{corollary}

The generality of our main reduction allows us to unlock techniques from the literature on leakage-resilient cryptography, specifically results related to the robustness of the LWE assumption~\cite{goldwasser2010robustness,brakerski2013classical,Mic18,BD20}, and go much further. In particular, using a variant of the reduction of \cite{Mic18} modified to handle $k$-sparse secrets (discussed further in Section~\ref{sec:techoverview}) we show  the following corollary.  In the corollaries, the condition $n\gg \ell \log_2 q$ (resp. $k\log_2(n/k) \gg \ell \log_2(q)$) comes from the entropy of random $\pm 1$ vectors (resp. random $k$-sparse vectors). 

\begin{corollary}[Informal Version of Corollary~\ref{cor:LWE-to-sparse-CLWE}]\label{corintro}
For 
\begin{align*}
\gamma &= O\left( \sqrt{k\cdot \log n} \right) \text{  and  } \beta = O \left(\frac{ \sigma \sqrt{k}}{q} \right),
\end{align*}
we have a polynomial (in $n$) time reduction from standard decisional LWE, in dimension $\ell$, with $n$ samples, modulus $q$, and noise parameter $\sigma$, to decisional CLWE in dimension $n$ {\em with $k$-sparse norm-$1$ secrets} and parameters $\gamma$ and $\beta$, as long as $k \log_2(n/k) \gg \ell \log_2(q)$ and $\sigma \gg 1$.
\end{corollary}

Looking ahead, we note that Corollary~\ref{corintro} will help us derive improved hardness for the problem of learning mixtures of Gaussians. Towards that end, it is worth stepping back and examining how far one can push Corollary~\ref{corintro}.
The LWE problem is believed to be exponentially hard; that is, in $\ell$ dimensions with a modulus $q = \mathsf{poly}(\ell)$ and error parameter $\sigma = \mathsf{poly}(\ell)$, LWE is believed to be hard for algorithms that run in  $2^{\ell^\epsilon}$ time using $m = 2^{\ell^{\epsilon}}$ samples, for any $\epsilon < 1$ (see, e.g. \cite{PeikertLindner}). Breaking this sub-exponential barrier not only has wide-ranging consequences for lattice-based cryptography, but also to the ongoing NIST post-quantum standardization competition~\cite{NIST} where better algorithms for LWE will lead NIST to reconsider the current parameterization of LWE-based encryption and signature schemes. 

Assuming such a sub-exponential hardness of LWE, we get the hardness of CLWE with
$$\gamma = (\log n)^{\frac{1}{2} + \delta}\log \log n$$
for an arbitrarily small constant $\delta=\delta(\epsilon)$. On the other hand, under a far more conservative polynomial-hardness assumption on LWE, we get the hardness of CLWE with 
$\gamma = n^{\delta}$ for an arbitrarily small $\delta > 0$.

Combining our main reduction with the known classical reduction from worst-case lattice problems to LWE~\cite{brakerski2013classical} gives us classical worst-case hardness of CLWE.

\begin{corollary}[Classical Worst-case Hardness of CLWE, informal]\label{cor:classical-hardness-clwe}
There is an efficient classical reduction from worst-case $\mathsf{poly}(n/\beta)$-approximate $\mathsf{gapSVP}$ in $\sqrt{n}$ dimensions, to decisional CLWE in $n$ dimensions with $\gamma =  \widetilde{\Omega}(\sqrt{n})$ and arbitrary $\beta = 1/\mathsf{poly}(n)$.
\end{corollary}

Finally, in Appendix~\ref{appendix:clwe-to-lwe}, we also show a reduction in the opposite direction, that is, from (discrete-secret) CLWE to LWE. Modulo the discrete secret requirement, this nearly completes the picture of the relationship between LWE and CLWE.  In turn, our reverse reduction can be combined with the other theorems in this paper to show a search-to-decision reduction for (discrete-secret) CLWE.

\subsection{Improved Hardness of Learning Mixtures of Gaussians} 

Bruna, Regev, Song and Tang~\cite{CLWE} used the hardness of CLWE to deduce hardness of problems in machine learning, most prominently the hardness of learning mixtures of Gaussians. We use our improved hardness result for CLWE to show improved hardness results for learning mixtures of Gaussians. First, let us start by describing the problem of Gaussian mixture learning.

\paragraph{Background on Gaussian Mixture Learning} 
The problem of learning a mixture of Gaussians is of fundamental importance in many fields of science~\cite{titterington1985statistical,MPbook}. Given a set of $g$ multivariate Gaussians in $n$ dimensions, parameterized by their means $\bmu_i \in \R^n$, covariance matrices $\Sigma_i \in \R^{n\times n}$, and non-negative weights $w_1,\ldots,w_g$ summing to one, the Gaussian mixture model is defined to be the distribution generated by picking a Gaussian $i\in [g]$ with probability $w_i$ and outputting a sample from $\Norm(\bmu_i,\Sigma_i)$.

Dasgupta~\cite{Dasgupta99} initiated the study of this problem in computer science. 
A strong notion of learning mixtures of Gaussians is that of {\em parameter estimation}, \emph{i.e.} to estimate all $\bmu_i$,  $\Sigma_i$ and $w_i$ given samples from the distribution. If one assumes the Gaussians in the mixture are well-separated, then the problem is known to be tractable for a constant number of Gaussian components \cite{Dasgupta99, sanjeev2001learning, vempala2002spectral, achlioptas2005spectral, kannan2005spectral, dasgupta2007probabilistic,
brubaker2008isotropic, KMV10, moitra2010settling, belkin2015polynomial, hardt2015tight, regev2017learning, hopkins2018mixture, kothari2018robust, diakonikolas2018list}. 
Moitra and Valiant \cite{moitra2010settling} and Hardt and Price \cite{hardt2015tight} also show that for parameter estimation, there is an information theoretic sample-complexity lower bound of $(1/\gamma)^g$ where $\gamma$ is the separation parameter and $g$ the number of Gaussian components.

Consequently, it makes sense to ask for a weaker notion of learning, namely {\em density estimation}, where, given samples from the Gaussian mixture, the goal is to output a ``density oracle'' (e.g. a circuit) that on any input $\vecx \in \R^n$, outputs an estimate of the density at $\vecx$~\cite{feldman2006pac}. The statistical distance between the density estimate and the true density must be at most a parameter $0 \leq \epsilon \leq 1$. The sample complexity of density estimation does not suffer from the exponential dependence in $g$, as was the case for parameter estimation. In fact, Diakonikolas, Kane, and Stewart \cite{diakonikolas2017statistical} show a $\poly(n, g, 1/\epsilon)$ upper bound on the information-theoretic sample complexity, by giving an {\em exponential-time} algorithm. 

Density estimation seems to exhibit a statistical-computational trade-off. While \cite{diakonikolas2017statistical} shows a polynomial upper bound on sample complexity, all known algorithms for density estimation, e.g., \cite{moitra2010settling}, run in time $(n/\epsilon)^{f(g)}$ for some $f(g) \ge g$. This is polynomial-time only for constant $g$. 
Furthermore, \cite{diakonikolas2017statistical} shows that even density estimation of Gaussian mixtures incurs a super-polynomial lower bound in the restricted statistical query (SQ) model~\cite{kearns1998efficient, feldman2017statistical}. Explicitly, they show that any SQ algorithm giving density estimates requires $n^{\Omega(g)}$ queries to an SQ oracle of precision $n^{-O(g)}$; this is super-polynomial as long as $g$ is super-constant. However, this lower bound does not say anything about arbitrary polynomial time algorithms for density estimation. 

The first evidence of computational hardness of density estimation for Gaussian mixtures came from the work of Bruna, Regev, Song and Tang~\cite{CLWE}. They show that being able to output a density estimate for mixtures of $g = \Omega(\sqrt{n})$ Gaussians implies a quantum polynomial-time algorithm for worst-case lattice problems. This leaves a gap between $g = O(1)$ Gaussians, which is known to be learnable in polynomial time, versus $g = \Omega(\sqrt{n})$ Gaussians, which is hard to learn. What is the true answer?

\begin{figure*}
\centering
\begin{tabular}{ |p{2cm}||p{3.8cm}|p{4.6cm}|p{2.1cm}|p{1.8cm}|}
 \hline
 \multicolumn{5}{|c|}{Summary of GMM Hardness Results} \\
 \hline
 & LWE Assumption  & \multirow{2}{*}{Gaussian Components} & \multirow{2}{*}{Run-time} & \multirow{2}{*}{Samples} \\
 & (samples, time, adv.) & & & \\
 \hline
 Corollary~\ref{cor:main-result-poly-lwe}        & $\left(\ell^{1/\epsilon}, \poly(\ell), \frac{1}{\poly(\ell)} \right)$          & $O\left(n^{\epsilon/2} \cdot \log n \right)$  & $n^{\omega(1)}$ & $\poly(n)$ \\
 Corollary~\ref{cor:main-result}                 & $\left(2^{\ell^{\delta}}, 2^{O(\ell^{\epsilon})}, \frac{1}{2^{O \left(\ell^{\delta} \right)}} \right)$          & $O\left( (\log n)^{\frac{1}{2} + \frac{1}{2\delta}} \cdot \sqrt{\log \log n} \right) $  & $\Omega\left( 2^{(\log n)^{\epsilon/\delta}} \right)$   & $\poly(n)$ \\
 Corollary~\ref{cor:main-result}                 & $\left( 2^{\ell^{\delta}}, 2^{O(\ell^{\epsilon})}, \frac{1}{\poly(\ell)} \right)$          & $O\left((\log n)^{\frac{1}{2\delta}} \cdot \log \log n\right) $  & $\Omega\left( 2^{(\log n)^{\epsilon/\delta}} \right)$  & $\poly(\log n)$ \\
 \hline
\end{tabular}
\caption{This tables summarizes our hardness results for density estimation of GMM. Throughout, $\delta, \epsilon \in (0,1)$ are arbitrary constants with $\delta < \epsilon$, $\ell$ is the dimension of LWE, and the Gaussians live in $\R^n$. ``Adv.'' stands for the advantage of the LWE distinguisher. As an example, the first row says for an arbitrary constant $0 < \epsilon < 1$, assuming standard, decisional LWE has no solver in dimension $\ell$ with $1/\poly(\ell)$ advantage given $\ell^{1/\epsilon}$ samples and $\mathsf{poly}(\ell)$ time, then any algorithm solving GMM density estimation given access to $\poly(n)$ samples from an arbitrary Gaussian mixture with at most $O(n^{\epsilon/2} \cdot \log n)$ Gaussian components must take super-polynomial in $n$ time.
}
\label{fig:gmm-summary}
\end{figure*}

\paragraph{Our Results on the Hardness of Gaussian Mixture Learning} Armed with our reduction from LWE to CLWE, and leakage-resilience theorems from the literature which imply Corollaries~\ref{cor1} and \ref{corintro}, we demonstrate a rich landscape of lower-bounds for density estimation of Gaussian mixtures. 

Using Corollary~\ref{cor1}, we show a hardness result for density estimation of Gaussian mixtures that improves on \cite{CLWE} in two respects. First, we show hardness of density estimation for $g = n^{\epsilon}$ Gaussians in $n$ dimensions for any $\epsilon>0$, assuming the polynomial-time hardness of LWE. Combined with the quantum reduction from worst-case lattice problems to LWE~\cite{regev2009lattices}, this gives us hardness for $n^{\epsilon}$ Gaussians under the quantum worst-case hardness of lattice problems. This improves on \cite{CLWE} who show hardness for $\Omega(\sqrt{n})$ Gaussians under the same assumption.
Secondly, our hardness of density estimation can be based on the {\em classical} hardness of lattice problems. 

The simplicity and generality of our main reduction from LWE to CLWE gives us much more. For one, assuming the sub-exponential hardness of LWE, we show that density estimation of $g = (\log n)^{1 + \epsilon}$ Gaussians cannot be done in polynomial time given a polynomial number of samples (where $\epsilon>0$ is an arbitrarily small  constant). This brings us very close to the true answer: we know that $g = O(1)$ Gaussians can be learned in polynomial time; whereas $g = (\log n)^{1 + \epsilon}$ Gaussians cannot, under a standard assumption in lattice-based cryptography (indeed, one that underlies post-quantum cryptosystems that are about to be standardized by NIST~\cite{NIST}).

We can stretch this even a little further. We show the hardness of density estimation for $g = (\log n)^{1/2 + \epsilon}$ Gaussians given $\mathsf{poly}(\log n)$ samples (where $\epsilon > 0$ is an arbitrary constant). This may come across as a surprise: is the problem even solvable information-theoretically given such few samples? It turns out that the sample complexity of density estimation for our hard instance, and also the hard instance of \cite{diakonikolas2017statistical}, is poly-logarithmic in $n$. In fact, we show (in Corollary~\ref{cor:gmm-upper-bound}) a quasi-polynomial time algorithm that does density estimation for our hard instance with $(\log n)^{1 + 2 \epsilon}$ samples. In other words, this gives us a tight computational gap for density estimation for the Gaussian mixture instances we consider.

These results are summarized below and more 
succinctly in Figure~\ref{fig:gmm-summary}. The reader is referred to  Section~\ref{sec:gmm-hardness} for the formal proofs.

\begin{theorem}[Informal Version of Corollary~\ref{cor:main-result} and Corollary~\ref{cor:main-result-poly-lwe}]\label{thm:main-thm-gmm-intro}
We give the following lower bounds for GMM density estimation based on LWE assumptions of varying strength.
\begin{enumerate}
    \item\label{item:poly-n} Assuming standard polynomial hardness of LWE, any density estimator for $\R^n$ that can solve arbitrary mixtures with at most $n^{\epsilon}$ Gaussian components, given $\poly(n)$ samples from the mixture, requires super-polynomial time in $n$ for arbitrary constant $\epsilon > 0$.
    \item\label{item:log-n} For constant $\epsilon \in (0,1)$, assuming $\ell$-dimensional $\LWE$ is hard to distinguish with advantage $1/2^{\ell^{\epsilon}}$ in time $2^{\ell^{\epsilon}}$, any density estimator for $\R^n$ that can solve arbitrary mixtures with at most roughly $(\log n)^{\frac{1}{2} + \frac{1}{2\epsilon}}$ Gaussian components, given $\poly(n)$ samples from the mixture, requires super-polynomial in $n$.
    \item\label{item:root-log-n} For constant $\epsilon \in (0,1)$, assuming $\ell$-dimensional $\LWE$ is hard to distinguish with advantage $1/\poly(\ell)$ in time $2^{\ell^{\epsilon}}$, any density estimator for $\R^n$ that can solve arbitrary mixtures with at most roughly $(\log n)^{\frac{1}{2 \epsilon}}$ Gaussian components, given $\poly(\log n)$ samples from the mixture, requires super-polynomial in $n$ time.
\end{enumerate}
\end{theorem}

\subsection{Other Applications}

Recent results have shown reductions from CLWE to other learning tasks as well, including learning a single periodic neuron~\cite{SZB21}, detecting backdoors in certain models~\cite{GKVZ22}, and improperly learning halfspaces in various error models~\cite{tiegel2022hardness,diakonikolas2022cryptographic}.\footnote{More precisely, Diakonikolas, Kane, Manurangsi and Ren~\cite{diakonikolas2022cryptographic} use our techniques to reduce from LWE instead of CLWE.} Our main result allows these results to be based on the hardness of LWE instead of CLWE.

In fact, we mention that our reduction can be used to show further hardness of the above learning tasks. For example, Song, Zadik and Bruna \cite{SZB21} directly show CLWE-hardness of learning single periodic neurons, \ie, neural networks with no hidden layers and a periodic activation function  $\varphi(t) = \cos(2 \pi \gamma t)$ with frequency $\gamma$. Our reduction from LWE to CLWE shows that this hardness result can be based directly on LWE instead of worst-case lattice assumptions, as done in \cite{CLWE}. Furthermore, our results expand the scope of their reduction in two ways:
\begin{enumerate}
    \item Their reduction shows hardness of learning periodic neurons with frequency $\gamma \geq \sqrt{n}$, while ours, based on exponential hardness of LWE, applies to frequencies almost as small as $\gamma = \log n$, which covers a larger class of periodic neurons. 
    
    \item Second, the hardness of $k$-sparse CLWE from (standard) LWE shows that even learning sparse features (instead of features drawn from the unit sphere $S^{n-1}$) is hard under LWE for appropriate parameter settings.
\end{enumerate}
This flexibility in $\gamma$ and in the sparsity of the secret distribution translates similarly for the other learning tasks mentioned, namely detecting backdoors in certain models~\cite{GKVZ22} and improperly learning halfspaces in various error models \cite{tiegel2022hardness,diakonikolas2022cryptographic}. For hardness of detecting backdoors~\cite{GKVZ22}, this flexibility means reducing the magnitude of undetectable backdoor perturbations (in $\ell_2$ and $\ell_0$ norms). For hardness of learning halfspaces, this flexibility means that agnostically learning noisy halfspaces is hard even if the optimal halfspace is now sparse.\footnote{The Veronese map translates a $k$-sparse degree-$d$ polynomial threshold function in dimension $n$ to a $\binom{k+d}{d}$-sparse linear threshold function (\ie, halfspace) in dimension $\binom{n+d}{d}$.}

\subsection{Perspectives and Future Directions}

The main technical contribution of our paper is a reduction from the learning with errors (LWE) problem to its continuous analog, CLWE. A powerful outcome of our reduction is the fact that one can now bring to bear powerful tools from the study of the LWE problem to the study of continuous LWE and its downstream applications. We show two such examples in this paper: the first is a classical worst-case to average-case reduction from the approximate shortest vector problem on lattices to continuous LWE; and the second is an improved hardness result for the well-studied problem of learning mixtures of Gaussians. 
We believe much more is in store.

For one, while we show a search-to-decision reduction for discrete-secret CLWE (see  Appendix~\ref{appendix:clwe-to-lwe}), we still do not know such a reduction for general CLWE. This is in contrast to multiple search-to-decision reductions of varying complexity and generality for the LWE problem~\cite{regev2009lattices,MM11}. Secondly, while there has been some initial exploration of the cryptographic applications of the continuous LWE problem~\cite{BNHR22}, constructing {\em qualitatively new} cryptographic primitives or {\em qualitatively better} cryptographic constructions is an exciting research direction. A recent example is the result of \cite{GKVZ22} who show use the hardness of CLWE to undetectably backdoor neural networks.

Finally, in terms of the hardness of learning mixtures of Gaussians, the question remains: what is the true answer? The best algorithms for learning mixtures of Gaussians~\cite{moitra2010settling} run in polynomial time only for a constant number of Gaussians. We show hardness (under a plausible setting of LWE) for roughly $\sqrt{\log n}$ Gaussians. 

In our hard instance, the Gaussian components live on a line, and indeed a one-dimensional lattice. For such Gaussians, we know from Bruna et al. \cite{CLWE} that there exists an algorithm running in time roughly $2^{O(g^2)}$, which becomes almost polynomial at the extremes of our parameter settings. Thus, we show the best lower bound possible for our hard instance. (In fact, for our hard instance, we can afford to enumerate over all sparse secret directions to get a solver with a similar run-time as \cite{CLWE} but with much smaller sample complexity. See Corollary~\ref{thm:algorithm-brute-force-hclwe} for details.)

There remain three possibilities: 
\begin{itemize} 
\item There is a different hard instance for learning any super-constant number of Gaussians in polynomial time, and hardness can be shown by reduction from lattice problems; or
\item There is a different hard instance for learning any super-constant number of Gaussians in polynomial time, but lattice problems are not the source of hardness; or 
\item We live in algorithmica, where the true complexity of Gaussian mixture learning is better than $n^{f(g)}$ and looks perhaps more like $\mathsf{poly}(n)\cdot 2^{g^2}$, despite what SQ lower bounds suggest~\cite{diakonikolas2017statistical}. 
\end{itemize} 
If we believe in the first two possibilities, a natural place to look for a {\em different} hard instance is \cite{diakonikolas2017statistical}, who consider a family of $g$ Gaussian pancakes centered at the roots of a Hermite polynomial. This allows them to match the first $2g-1$ moments with that of the standard Gaussian. A tantalizing open problem is to try and prove hardness for their distribution for all algorithms, not just SQ algorithms, possibly under some cryptographic assumptions or perhaps even lattice assumptions.

\section{Technical Overview}
\label{sec:techoverview}

\begin{figure}
\centering
\begin{tabular}{ |l||l|l|l|}
 \hline
 \multicolumn{4}{|c|}{Reducing Fixed-Norm LWE to CLWE (Theorem~\ref{thm:fixed-norm-lwe-clwe})} \\
 \hline
 & Samples & Secrets & Errors\\
 \hline
 Fixed-Norm LWE                                         & $U(\Z_q^{n})$      & $\S$           & $D_{\Z, \sigma_1}$ \\
 Step 1 (Lemma \ref{lemma:discrete-to-continuous-errors})  & $U(\Z_q^{n})$      & $\S$                & $D_{\sigma_2}$     \\
 Step 2 (Lemma \ref{lemma:discrete-to-continuous-samples}) & $U(\T_{q}^{n})$    & $\S$                & $D_{\sigma_3}$     \\
 CLWE (Lemma \ref{lemma:uniform-to-gaussian-samples})    & $D_1^{n}$       & $\frac{1}{r} \cdot \S$                    & $D_{\beta}$   \\
 \hline
\end{tabular}
\caption{This table shows the steps in the reduction from fixed-norm LWE to CLWE (with discrete secret distribution $\frac{1}{r}\cdot \S$ of unit norm; to reduce to continuous uniform unit-vector secrets, one can apply Lemma~\ref{lemma:wc-to-ac-clwe}). All of the reductions in the table are sample preserving, dimension preserving, and advantage preserving (up to $\negl(\secp)$ additive loss). To reduce from LWE with secrets $\vecs \sim U(\Z_q^n)$ (instead of a fixed-norm distribution), we first apply Theorem~\ref{thm:uniform-to-binary-secrets} and then we perform the steps above.} 
\label{fig:steps-copy}
\end{figure}

\subsection{From Fixed-Norm LWE to CLWE} 
The goal of our main theorem 
(Theorem~\ref{thm:main-lwe-clwe-reduction-informal}) is to reduce from the fixed-norm LWE problem to CLWE. This involves a number of transformations, succinctly summarized in Figure~\ref{fig:steps-copy}. Given samples $(\mathbf{a},b=\langle \mathbf{a},\mathbf{s}\rangle + e \pmod{q}) \in \Zq^{n+1}$, we do the following:

\begin{enumerate} 
\item  First, we turn the errors (in $b$) from discrete to continuous Gaussians by adding a small continuous Gaussian to the LWE samples, using the smoothing lemma~\cite{micciancio2007worst}.
\item  Secondly, we turn the samples $\mathbf{a}$ from discrete to continuously uniform over the torus by doing the same thing, namely adding a continuous Gaussian noise, and once again invoking appropriate smoothing lemmas from \cite{regev2009lattices,micciancio2007worst}.
\item Third, we go from uniform samples $\mathbf{a}$ to Gaussian samples. Boneh, Lewi, Montgomery and Raghunathan ~\cite{BLMR13} give a general reduction from $U(\Z_q^n)$ samples to ``coset-sampleable'' distributions, and as one example, they show how to reduce discrete uniform samples to discrete Gaussian samples, at the cost of a $\log q$ multiplicative overhead in the dimension, which is  unavoidable information-theoretically. We improve this reduction and circumvent this lower bound in the continuous version by having \emph{no overhead} in the dimension, \emph{i.e.} the dimension of both samples are the same. The key ingredient to this improvement is a simple Gaussian pre-image sampling algorithm, which on input $z \sim U([0,1))$, outputs $y$ such that $y = z \pmod{1}$ and $y$ is statistically close to a continuous Gaussian (when marginalized over $z \sim U([0,1))$). (See Lemma \ref{lemma:gaussian-preimage-sampling} for a more precise statement.)
\item This finishes up our reduction! The final thing to do is to scale down the secret and randomly rotate it to ensure that it is a uniformly random unit vector.
\end{enumerate} 

We note that up until the final scaling down and re-randomization step, our reduction is {\em secret-preserving}.

\subsection{Hardness of Gaussian Mixture Learning}
Bruna et al. \cite{CLWE} show that a homogeneous version of CLWE, called hCLWE, has a natural interpretation as a certain distribution of mixtures of Gaussians. They show that any distinguisher between the hCLWE distribution and the standard multivariate Gaussian is enough to solve CLWE. Therefore, an algorithm for density estimation for Gaussian mixtures, which is a harder problem than distinguishing between that mixture and the standard Gaussian, implies a solver for CLWE. The condition that $g > \sqrt{n}$ is a consequence of their reduction from worst-case lattice problems.

Our {\em direct} reduction from LWE to CLWE opens up a large toolkit of techniques that were developed in LWE-based cryptography. In this work, we leverage tools from leakage-resilient cryptography~\cite{brakerski2013classical,Mic18,BD20} to improve and generalize the hard instance of \cite{CLWE}. The key observation is that the number of Gaussians $g$ in the mixture at the end of the day roughly corresponds to the norm of the secrets in LWE. Thus, the hardness of LWE with low-norm secrets will give us the hardness of Gaussian mixture learning with a small number of Gaussians.

Indeed, we achieve this by reducing LWE to $k$-sparse LWE. We call a vector $\vecs \in \{+1, 0, -1\}^n$ $k$-sparse if it has exactly $k$ non-zero entries. We show the following result:

\begin{theorem}[Informal Version of Corollary~\ref{cor:useful-lwe-to-k-sparse-lwe}]\label{thm:main-k-sparse}
Assume LWE in dimension $\ell$ with $n$ samples is hard with secrets $\vecs \sim \Z_q^\ell$ and errors of width $\sigma$. Then, LWE in dimension $n$ with $k$-sparse secrets is hard for errors of width $O(\sqrt{k} \cdot \sigma)$, as long as $k \log_2(n/k) \gg \ell \log_2(q)$.
\end{theorem}

It turns out that for our purposes, the quantitative tightness of our theorem is important. Namely, we require that the blowup in the noise depends polynomially only on $k$ and not on other parameters. 
Roughly speaking, the reason is that if we have a blow-up factor of $r$, for our LWE assumption, we need $q/\sigma \gg r$ for the resulting CLWE distribution to be meaningful. For our parameter settings, if $r$ depends polynomially on the dimension $n$ (the dimension of the ambient space for the Gaussians) or the number of samples $m$, then we require sub-exponentially large modulus-to-noise ratio in our LWE assumption, which is a notably stronger assumption.
Indeed, the noise blow-up factor of the reduction we achieve and use is $O(\sqrt{k})$.

Our proof of this theorem uses a variant of the proof of \cite{Mic18} to work with $k$-sparse secrets.\footnote{The techniques of Brakerski et al.~\cite{brakerski2013classical}, who show the hardness of binary secret LWE, can also be easily modified to prove $k$-sparse hardness, but the overall reduction is somewhat more complex. For this reason, we choose to show how to modify the reduction of \cite{Mic18}.} We note that Brakerski and D{\"{o}}ttling \cite{BD20} give a general reduction from LWE to LWE with arbitrary secret distributions with large enough entropy, but the noise blowup when applying their results directly to $k$-sparse secrets is roughly $\sqrt{k m n} = k^{\omega(1)}$ for parameter settings we consider.

For a full description of the proof of Theorem~\ref{thm:main-k-sparse}, the reader is referred to Section~\ref{sec:k-sparse-lwe-hardness}. 

%% file: prelims.tex
\section{Preliminaries}
\label{sec:prelims}

For a distribution $\mathcal{D}$, we write $x \sim \mathcal{D}$ to denote a random variable $x$ being sampled from $\mathcal{D}$. For any $n \in \N$, we let $\mathcal{D}^n$ denote the $n$-fold product distribution, i.e. $(x_1, \dots, x_n) \sim \mathcal{D}^n$ is generated by sampling $x_i \sim_{\text{i.i.d.}} \mathcal{D}$ independently. For any finite set $S$, we write $U(S)$ to denote the discrete uniform distribution over $S$; we abuse notation and write $x \sim S$ to denote $x \sim U(S)$. For any continuous set $S$, we write $U(S)$ to denote the continuous uniform distribution over $S$ (i.e. having support $S$ and constant density); we also abuse notation and write $x \sim S$ to denote $x \sim U(S)$.

For distributions $\cD_1, \cD_2$ supported on a measurable set $\mathcal{X}$, we define the statistical distance between $\cD_1$ and $\cD_2$ to be $\Delta(\cD_1, \cD_2) = \frac{1}{2} \int_{x \in \mathcal{X}} | \cD_1(x) - \cD_2(x)| dx$. We say that distributions $\cD_1, \cD_2$ are $\epsilon$-close if $\Delta(\cD_1, \cD_2) \leq \epsilon$. For a distinguisher $\mathcal{A}$ running on two distributions $\cD_1$, $\cD_2$, we say that $\mathcal{A}$ has advantage $\epsilon$ if
\[ \left| \Pr_{x \sim \cD_1}[\mathcal{A}(x) = 1]-  \Pr_{x \sim \cD_2}[\mathcal{A}(x) = 1] \right| = \epsilon, \]
where the probability is also over any internal randomness of $\mathcal{A}$.

We let $I_{n \times n} \in \{0,1\}^{n \times n}$ denote the $n \times n$ identity matrix. When $n$ is clear from context, we write this simply as $I$. For any matrix $M \in \R^{m \times n}$, we let $M^\top$ be its transpose matrix, and for $\ell \in [n]$, we write $M_{[\ell]} \in \R^{m \times \ell}$ to denote the submatrix of $M$ consisting of just the first $\ell$ columns, and we write $M_{]\ell[} \in \R^{m \times (n - \ell)}$ to denote the submatrix of $M$ consisting of all but the first $\ell$ columns.

For any vector $\vecv \in \R^n$, we write $\norm{\vecv}$ to mean the standard $\ell_2$-norm of $\vecv$, and we write $\norm{\vecv}_{\infty}$ to denote the $\ell_{\infty}$-norm of $\vecv$, meaning the maximum absolute value of any component. For $n \in \N$, we let $S^{n-1} \subset \R^n$ denote the $(n-1)$-dimensional sphere embedded in $\R^n$, or equivalently the set of unit vectors in $\R^n$. By $\Z_q$, we refer to the ring of integers modulo $q$, represented by $\{0, \dots, q-1\}$. By $\T_q$, we refer to the set $\R/q\Z = [0,q) \subseteq \R$ where addition (and subtraction) is taken modulo $q$ (i.e. $\T_q$ is the torus scaled up by $q$). We denote $\T := \T_1$ to be the standard torus. By taking a real number mod $q$, we refer to taking its representative as an element of $\T_q$ in $[0,q)$ unless stated otherwise.

\begin{definition}[Min-Entropy]
For a discrete distribution $\cD$ with support $S$, we let $\minentropy(\cD)$ denote the min-entropy of $\cD$,
\[ \minentropy(\cD) = - \log_2 \left( \max_{s \in S} \Pr_{x \sim \cD} [x = s] \right). \]
\end{definition}

\begin{lemma}[Leftover Hash Lemma \cite{haastad1999pseudorandom}]\label{lemma:leftover-hash-lemma}
Let $\ell, n, q \in \N, \epsilon \in \R_{>0}$, and let $\S$ be a distribution over $\{-1,0,1\}^n \subseteq \Z_q^n$. Suppose $\minentropy(\S) \geq \ell \log_2(q) + 2 \log_2(1/\epsilon)$. Then, the distributions given by $(A, A \vecs \pmod{q})$ and $(A, \vecb)$ where $A \sim \Z_q^{\ell \times n}$, $\vecs \sim \S$, $\vecb \sim \Z_q^\ell$  have statistical distance at most $\epsilon$.
\end{lemma}

\subsection{Lattices and Discrete Gaussians}

A rank $n$ integer lattice is a set $\Lambda = \matB \Z^n \subseteq \Z^d$ of all integer linear combinations of $n$ linearly independent vectors $\matB = [\vecb_1, \ldots, \vecb_n]$ in $\Z^d$. The dual lattice $\Lambda^*$ of a lattice $\Lambda$ is defined as the set of all vectors $\vecy \in \R^d$ such that $\inner{\vecx, \vecy} \in \Z$ for all $\vecx \in \Lambda$.

For arbitrary $\vecx \in \R^n$ and $\vecc \in \R^n$, we define the Gaussian function
\[ \rho_{s, \vecc}(\vecx) = \exp \left(- \pi \norm{(\vecx - \vecc)/s}^2 \right). \]
Let $D_{s, \vecc}$ be the corresponding distribution with density at $\vecx \in \R^n$ given by $\rho_{s, \vecc}(x) / s^n$, namely the $n$-dimensional Gaussian distribution with mean $\vecc$ and covariance matrix $s^2/(2\pi) \cdot I_{n \times n}$. When $\vecc = 0$, we omit the subscript notation of $\vecc$ on $\rho$ and $D$. 

For an $n$-dimensional lattice $\Lambda \subseteq \R^n$ and point $\vecc \in \R^n$, we can define the \emph{discrete Gaussian of width $s$} to be given by the mass function
\[ D_{\Lambda + \vecc, s}(\vecx) = \frac{\rho_s(\vecx)}{\rho_s(\Lambda + \vecc)} \]
supported on $\vecx \in \Lambda + \vecc$, where by $\rho_s(\Lambda + \vecc)$ we mean $\sum_{\vecy \in \Lambda} \rho_s(\vecy + \vecc)$.

We now give the smoothing parameter as defined by \cite{regev2009lattices} and some of its standard properties.

\begin{definition}[\cite{regev2009lattices}, Definition 2.10]
For an $n$-dimensional lattice $\Lambda$ and $\epsilon > 0$, we define $\eta_{\epsilon}(\Lambda)$ to be the smallest $s$ such that $\rho_{1/s}(\Lambda^* \setminus \{\mathbf{0}\}) \leq \epsilon$.
\end{definition}

\begin{lemma}[\cite{regev2009lattices}, Lemma 2.12]\label{lemma:smoothing-parameter-estimate}
For an $n$-dimensional lattice $\Lambda$ and $\epsilon > 0$, we have
\[ \eta_{\epsilon}(\Lambda) \leq \sqrt{\frac{\ln(2n(1 + 1/ \epsilon))}{\pi}} \cdot \lambda_n(\Lambda). \]
Here $\lambda_i(\Lambda)$ is defined as the minimum length of the longest vector in a set of $i$ linearly independent vectors in $\Lambda$.
\end{lemma}

\begin{lemma}[\cite{regev2009lattices}, Corollary 3.10]\label{lemma:discrete-plus-continuous}
For any $n$-dimensional lattice $\Lambda$ and $\epsilon \in (0, 1/2)$, $\sigma, \sigma' \in \R_{> 0}$, and $\vecz, \vecu \in \R^n$, if
\[ \eta_{\epsilon}(\Lambda) \leq \frac{1}{\sqrt{1/(\sigma')^2 + (\norm{\vecz}/\sigma)^2}},\]
then if $\vecv \sim D_{\Lambda + \vecu, \sigma'}$ and $e \sim D_{\sigma}$, then $\inner{\vecz, \vecv} + e$ has statistical distance at most $4 \epsilon$ from $D_{\sqrt{(\sigma' \norm{\vecz})^2 + \sigma^2}}$.
\end{lemma}

\begin{lemma}[\cite{micciancio2007worst}, Lemma 4.1]\label{lemma:gaussian-mod-lattice-looks-uniform}
For an $n$-dimensional lattice $\Lambda$, $\epsilon > 0$, $\vecc \in \R^n$ for all $s \geq \eta_{\epsilon}(\Lambda)$, we have
\[ \Delta(D_{s, \vecc} \mod P(\Lambda), U(P(\Lambda))) \leq \epsilon/2, \]
where $P(\Lambda)$ is the half-open fundamental parallelepiped of $\Lambda$.
\end{lemma}

\begin{lemma}[\cite{micciancio2007worst}, implicit in Lemma 4.4]\label{lemma:smoothing-swallows-shifting}
For an $n$-dimensional lattice $\Lambda$, for all $\epsilon > 0$, $\vecc \in \R^n$, and all $s \geq \eta_{\epsilon}(\Lambda)$, we have
\[ \rho_{s}(\Lambda + \vecc) = \rho_{s, -\vecc}(\Lambda) \in \left[ \frac{1 - \epsilon}{1 + \epsilon}, 1\right] \cdot \rho_s(\Lambda).\]
\end{lemma}

Now we recall other facts related to lattices.

\begin{lemma}[\cite{micciancio2013hardness}, Theorem 3]\label{lemma:discrete-gaussian-convolution}
Suppose $\vecv \in \Z^m$ with $\gcd(\vecv) = 1$, 
and suppose $y_i \sim D_{\Z, \sigma_i}^m$ for all $i \in [m]$. As long as $\sigma_i \geq \sqrt{2} \norm{\vecv}_{\infty}\eta_{\frac{\epsilon}{2m^2}}(\Z)$ for all $i \in [m]$, then we have $y = \sum_{i \in [m]} y_i v_i$ is $O(\epsilon)$-close to $D_{\Z, \sigma}$ where $\sigma = \sqrt{\sum_{i \in [m]} \sigma_i^2 v_i^2}$.
 \end{lemma}

\begin{lemma}[\cite{Mic18}, Lemma 2.2]\label{lemma:primitive-vector-negl}
For $\vecw \sim U(\Z_q^\ell)$, the probability that $\gcd(\vecw, q) \neq 1$ is at most $\log(q)/2^\ell$.
\end{lemma}

\begin{definition}
We say that a matrix $T \in \Z^{k \times m}$ is primitive if $T \Z^m = \Z^k$, i.e., if $T: \Z^m \rightarrow \Z^k$ is surjective.
\end{definition}

\begin{lemma}[\cite{Mic18}, Lemma 2.6]\label{lemma:mic18-gaussian-projections}
For any primitive matrix $T \in \Z^{k \times m}$ and positive reals $\alpha, \sigma > 0$, if $T T^\top = \alpha^2 I$ and $\sigma \ge \eta_\epsilon(\ker(T))$, then $T(D_{\Z^m, \sigma})$ and $D_{\Z^k, \alpha \sigma}$ are $O(\epsilon)$-close.
\end{lemma}

\subsection{Learning with Errors}

Throughout, we work with decisional versions of LWE, CLWE, and hCLWE.

\begin{definition}[LWE Distribution]\label{def:lwe-distribution}
Let $n, m, q \in \N$, let $\mathcal{A}$ be a distribution over $\R^{n}$, $\S$ be a distribution over $\Z^n$, and $\mathcal{E}$ be a distribution over $\R$. We define $\LWE(m, \mathcal{A}, \S, \mathcal{E})$ to be distribution given by sampling $\veca_1, \cdots, \veca_m \sim \mathcal{A}$, $\vecs \sim \S$, and $e_1, \cdots, e_m \sim \mathcal{E}$, and outputting $(\veca_i, \vecs^\top \veca_i + e_i \pmod{q})$ for all $i \in [m]$. We refer to $n$ as the dimension and $m$ as the number of samples. (The modulus $q$ is suppressed from notation for brevity as it will be clear from context.)

We also consider the case where $\S$ is a distribution over $\Z^{n \times j}$ and $\mathcal{E}$ is a distribution over $\R^j$. In this case, the ouput of each sample is $(\veca_i, S^\top \veca_i + \vece_i \pmod{q})$, where $S \sim \S$ and $\vece_i \sim \mathcal{E}$.
\end{definition} 

\begin{definition}[CLWE Distribution \cite{CLWE}]\label{def:clwe-distribution}
Let $n, m \in \N, \gamma, \beta \in \R$, and let $\mathcal{A}$ be a distribution over $\R^{n}$ and $\S$ be a distribution over $S^{n-1}$. Let $\CLWE(m, \mathcal{A}, \S, \gamma, \beta)$ be the distribution given by sampling $\veca_1, \cdots, \veca_m \sim \mathcal{A}$, $\vecs \sim \S, e_1, \cdots, e_m \sim D_{\beta}$ and outputting $(\veca_i, \gamma \cdot \inner{\veca_i, \vecs} + e_i \pmod{1})$ for all $i \in [m]$. Explicitly, for one sample, the density at $(\vecy, z) \in \R^{n} \times [0, 1)$ is proportional to
\[ \mathcal{A}(\vecy) \cdot \sum_{k \in \Z} \rho_{\beta}(z + k - \gamma \cdot \inner{\vecy, \vecs}) \]
for fixed secret $\vecs \sim \S$. We refer to $n$ as the dimension and $m$ as the number of samples. We omit $\S$ if $\S = U(S^{n-1})$, as is standard for CLWE.
\end{definition}

\begin{definition}[hCLWE Distribution \cite{CLWE}]\label{def:hclwe-distribution}
Let $n, m \in \N, \gamma, \beta \in \R$, and let $\mathcal{A}$ be a distribution over $\R^{n \times m}$ and $\S$ be a distribution over $S^{n-1}$. Let $\hCLWE(m, \mathcal{A}, \S, \gamma, \beta)$ be the the distribution $\CLWE(m, \mathcal{A}, \S, \gamma, \beta)$, but conditioned on the fact that for all samples second entries are $0 \pmod{1}$.

Explicitly, for one sample, the density at $\vecy \in \R^n$ is proportional to
\[ \mathcal{A}(\vecy) \cdot \sum_{k \in \Z} \rho_{\beta}(k - \gamma \cdot \inner{\vecy, \vecs}) \]
for fixed secret $\vecs \sim \S$. We refer to $n$ as the dimension and $m$ as the number of samples. We omit $\S$ if $\S = U(S^{n-1})$, as is standard for hCLWE.
\end{definition}
Note that the hCLWE distribution is itself a mixture of Gaussians. Explicitly, for a secret $\vecs \sim \S$, we can write the density of $\hCLWE(1, D_1, \vecs, \gamma, \beta)$ at point $\vecx \in \R^n$ as proportional to
\begin{equation}\label{eq:hclwe-mixture}
\rho(\vecx) \cdot \sum_{k \in \Z} \rho_\beta(k - \gamma \cdot \inner{\vecs, \vecx}) = \sum_{k \in \Z} \rho_{\sqrt{\beta^2 + \gamma^2}}(k) \cdot \rho(\pi_{\vecs^\bot}(\vecx)) \cdot \rho_{\beta / \sqrt{\beta^2 + \gamma^2}} \left( \inner{\vecs, \vecx} - \frac{\gamma}{\beta^2 + \gamma^2} k \right),
\end{equation}
where $\pi_{\vecs^\bot}(\vecx)$ denotes the projection onto the orthogonal complement of $\vecs$. Thus, we can view hCLWE samples as being drawn from a mixture of Gaussians of width $\beta / \sqrt{\beta^2 + \gamma^2} \approx \beta/\gamma$ in the secret direction, and width 1 in all other directions.

\begin{definition}[Truncated hCLWE Distribution \cite{CLWE}]\label{def:truncated-hclwe-distribution}
Let $n, m, g\in \N, \gamma, \beta \in \R$, and let $\S$ be a distribution over $S^{n-1}$. Let $\hCLWE^{(g)}(m, \S, \gamma, \beta)$ be the the distribution $\hCLWE(m, D_1^n, \S, \gamma, \beta)$, but restricted to the central $g$ Gaussians, where by central $g$ Gaussians, we mean the central $g$ Gaussians in writing hCLWE samples as a mixture of Gaussians, as in Eq. \ref{eq:hclwe-mixture}. Explicitly, for secret $\vecs \sim \S$, the density of one sample at a point $\vecx \in \R^n$ is proportional to
\begin{align}\label{eqn:truncated-hclwe-pdf}
\sum_{k = -\floor{g/2}}^{\floor{(g-1)/2}} \rho_{\sqrt{\beta^2 + \gamma^2}}(k) \cdot \rho(\pi_{\vecs^\bot}(\vecx)) \cdot \rho_{\beta / \sqrt{\beta^2 + \gamma^2}} \left( \inner{\vecs, \vecx} - \frac{\gamma}{\beta^2 + \gamma^2} k \right).
\end{align}
\end{definition}

\begin{definition}[Density Estimation for the Gaussian Mixture Model (Definition~5.1 of \cite{CLWE}]
We say that an algorithm \emph{solves GMM density estimation} in dimension $n$ with $m$ samples and up to $g$ Gaussians if, when given $m$ samples from an arbitrary mixture of at most $g$ Gaussian components in $\R^n$, the algorithm outputs some density function (as an evaluation oracle) that has statistical distance at most $10^{-3}$ from the true density function of the mixture, with probability at least $9/10$ (over the randomness of the samples and the internal randomness of the algorithm).
\end{definition}

The following theorem tells us that distinguishing a truncated version of the hCLWE Gaussian mixture from the standard Gaussian is enough to distinguish the original Gaussian mixture from the standard Gaussian. In particular, we can use density estimation to solve hCLWE since the truncated version has a finite number of Gaussians. 

\begin{theorem}[Proposition 5.2 of \cite{CLWE}]\label{thm:hclwe-to-gmm}
Let $n, m \in \N$, $\gamma, \beta \in \R_{>0}$ with $\beta < 1/32$ and $\gamma \geq 1$. Let $\S$ be a distribution over $S^{n-1}$. For sufficiently large $m$ and for $g = 2 \gamma \sqrt{\ln m / \pi}$, if there is an algorithm running in time $T$ that distinguishes $\hCLWE^{(2g+1)}(m, \S, \gamma, \beta)$ and $D_1^{n \times m}$ with constant advantage, then there is a time $T + \poly(n,m)$ algorithm distinguishing  $\hCLWE(m, D_1^{n}, \S, \gamma, \beta)$ and $D_1^{n \times m}$ with constant advantage. In particular, if there is an algorithm running in time $T$ that solves density estimation with in dimension $n$ with $m$ samples and $g$ Gaussians, then there is a time $T + \poly(n,m)$ algorithm distinguishing $\hCLWE(m, D_1^{n}, \S, \gamma, \beta)$ and $D_1^{n \times m}$ with advantage at least $1/2$.
\end{theorem}

We also use a lemma which says that if CLWE is hard, then so is hCLWE.

\begin{lemma}[Lemma 4.1 of \cite{CLWE}]\label{lemma:clwe-to-hclwe} Let $\delta \in (0,1)$ be an input parameter. There is a randomized $\poly(n, m_1, 1/\delta)$-time reduction that maps $m_1$ samples from $\CLWE(D_1^n, \vecs, \gamma, \beta)$ to $m_2 = \Omega(\delta m_1)$ samples from $\hCLWE(D_1^n, \vecs, \gamma, \sqrt{\beta^2 + \delta^2})$ and maps $m_1$ samples from $D_1^n \times U(\T_1)$ to $m_2$ samples from $D_1^n$, with failure probability at most $1/1000$.
\end{lemma}

%% file: ksparse.tex
\section{Hardness of \texorpdfstring{$k$}{k}-sparse LWE}\label{sec:k-sparse-lwe-hardness}

In this section, we modify the proof of \cite{Mic18} to reduce from standard decisional LWE to a version where secrets are sparse, in the sense that they have few non-zero entries. The main changes we make to \cite{Mic18} are that we slightly modify the gadget matrix $Q$ and the matrix $Z$ to handle sparse secrets (using its notation).

For completeness, we give a self-contained proof.

\begin{definition}
For $k, n \in \N$ with $k \leq n$, let $\S_{n, k}$ be the subset of vectors in $\{-1, 0, +1\}^n$ with exactly $k$ non-zero entries. We call $\vecs \in \Z^n$ \emph{$k$-sparse} if $\vecs \in \S_{n,k}$.
\end{definition}

\begin{lemma}\label{lemma:min-entropy-of-sparse-secrets}
It holds that $\minentropy(\S_{n,k}) \geq k \log_2(n/k)$.
\end{lemma}
\begin{proof}
Observe that $|\S_{n,k}| = \binom{n}{k} \cdot 2^k$. Using the bound $(n/k)^k \leq \binom{n}{k}$, we have
\begin{align*}
    \minentropy(\S_{n,k}) \geq \log_2 \left( \left( 2 \cdot \frac{n}{k} \right)^k \right) \geq k \log_2 (n/k),
\end{align*}
as desired.
\end{proof}

Our main theorem in this section is the following:
\begin{theorem}\label{thm:regular-lwe-to-k-sparse-ours}
\sloppy Let $q,m,n,\ell, k \in \N$ with $1 < k < n$, and let $\sigma, \epsilon\in \R_{>0}$. Suppose $\log(q)/2^\ell = \negl(\secp), \sigma \geq 4\sqrt{\omega(\log \secp) + \ln n + \ln m}$, and $k \log(n/k) \geq (\ell + 1) \log_2(q) + \omega(\log \secp)$. Suppose there is no $T + \poly(n,m, \log(q), \log(\secp))$ time distinguisher with advantage $\epsilon - \negl(\secp)$ between $\LWE(n-1, \Z_q^\ell, \Z_q^{m \times \ell}, D_{\Z^m, \sigma})$ and $U(\Z_q^{\ell \times (n-1)} \times \Z_q^{m \times (n-1)})$, and further suppose there is no $T + \poly(n,m, \log(q), \log(\secp))$ time distinguisher with advantage $\epsilon - \negl(\secp)$ between $\LWE(n+1, \Z_q^{\ell + 1}, \Z_q^{m \times (\ell + 1)}, D_{\Z^m, 2\sigma})$ and $U(\Z_q^{(\ell +1) \times (n+1)} \times \Z_q^{m \times (n+1)})$. Then, there is no $T$ time distinguisher with advantage $2 \epsilon$ between $\LWE(m, \Z_q^n, \S_{n,k}, D_{\Z, \sigma'})$ and $U(\Z_q^{m \times n} \times \Z_q^m)$, where $\sigma' = 2 \sigma \sqrt{k+1}$.
\end{theorem}

\begin{definition}
Let $n,k \in \Z$ with $k \leq n$. For all $i \in [n]$, we define $\vece_i$ to be the $i$th standard basis column vector, i.e. having a 1 in the $i$th coordinate and 0s elsewhere. We then define $\vecu \in \Z^n$ to be $\vecu = \sum_{i=1}^k \vece_i$, i.e. 1s in the first $k$ coordinates and 0 elsewhere.
\end{definition}

\begin{lemma}\label{lemma:gadget-Q}
There is a $\poly(n)$-time computable matrix $Q \in \Z^{n \times (2n + 5)}$ such that $Q_{[n]}$ is invertible, $\vecu^\top Q_{[n]} = \vece_1^\top$, the vector $\vecv^\top = \vecu^\top Q_{]n[} \in \Z^{n+5}$ satisfies $\norm{\vecv}_2 = 2 \sqrt{k}$ and $\norm{\vecv}_{\infty} = 2$, and $Q_{]1[}(D_{\Z^{2n+4}, \sigma})$ and $D_{\Z^n, 2\sigma}$ are $\negl(\secp)/t$ close as long as $\sigma \geq \sqrt{6} \cdot \sqrt{\omega(\log \secp) + \ln n + \ln t}$ for a free parameter $t$.
\end{lemma}

\setcounter{MaxMatrixCols}{20}

\begin{proof}
We use essentially the same gadget $Q$ as in Lemma 2.7 of \cite{Mic18}, except we modify two entries of the matrix and add two columns. Specifically, we set $Q_{k, k+1} = 0$ (instead of $-1$), $Q_{k, n+k+1} = 0$ (instead of 1), and add two columns to the end that are all 0 except for two entries of $1$ in $Q_{k, 2n+4}$ and $Q_{k, 2n+5}$.

We will give it explicitly as follows. Let the matrix $X \in \Z^{n \times (n-1)}$ be defined by

\[ X = \begin{bmatrix}
-1 & & & & & & &\\
1 & -1 & & & & & &\\
& \ddots & \ddots & & & & &\\
& & 1 & -1 & & & &\\
& & & 1 & 0 & & &\\
& & & & 1 & -1 & &\\
& & & & & \ddots & \ddots &\\
& & & & & & 1 & -1\\
& & & & & & & 1\\
\end{bmatrix}, \]
where the row with the abnormal $0$ is the $k$th row. Similarly, let $Y \in \Z^{n \times (n-1)}$ be defined by
\[ Y = \begin{bmatrix}
1 & & & & & & &\\
1 & 1 & & & & & &\\
& \ddots & \ddots & & & & &\\
& & 1 & 1 & & & &\\
& & & 1 & 0 & & &\\
& & & & 1 & 1 & &\\
& & & & & \ddots & \ddots &\\
& & & & & & 1 & 1\\
& & & & & & & 1\\
\end{bmatrix}, \]
where the row with the abnormal $0$ is again the $k$th row. We then define $Q  \in \Z^{n \times (2n+5)}$ by
\[ Q = [\vece_1, X, -\vece_n, Y, \vece_n, \vece_1, \vece_1, \vece_k, \vece_k]. \]
First, notice that $Q_{[n]}$ is invertible, since it is upper-triangular with 1s on the diagonal. Next, notice that $\vecu^\top Q_{[n]} = \vece_1^\top$, as $\vecu^\top \vece_1 = 1$ and the sum of the first $k$ entries in each column of $X$ are all 0 by construction. We can write $\vecv^\top = \vecu^\top Q_{]n[} = [0, 2, 2, \cdots, 2, 0, \cdots, 0, 1,1, 1,1]$, which has $\ell_2$ norm
\[ \sqrt{(k-1) \cdot 2^2 + 4 \cdot 1^2} = 2\sqrt{k}. \]
It's clear to also see that $\norm{\vecv}_{\infty} = 2$. All that is remaining to show is that $Q_{]1[}(D_{\Z^{2n + 4}, \sigma})$ and $D_{\Z^n, 2 \sigma}$ are $\negl(\secp)/t$-close, which we do below.

To show that $Q_{]1[}(D_{\Z^{2n+4}, \sigma})$ and $D_{\Z^n, 2\sigma}$ are $\negl(\secp)/t$ close, we first prove the preconditions of, and then invoke, Lemma~\ref{lemma:mic18-gaussian-projections}.
Let $T = Q_{]1[} \in \Z^{n \times (2n + 4)}$. 

First, we show that $T$ is primitive. It suffices to show that for every standard basis column vector $\vece_i$, there is some $\vecg_i \in \Z^{2n+4}$ such that $\vece_i = T \vecg_i$.
For all $j \in [2n+4]$, we define $\vecf_j$ to be the $j$th standard basis column vector in $\R^{2n+4}$. 
Let $\vecg_1 = \vecf_{2n+1}$, and $\vecg_{k+1} = \vecf_k$. It can be easily checked that $\vece_1 = T \vecg_1$ and $\vece_{k+1} = T \vecg_{k+1}$. 
Then, for all $i$ such that $1 < i \le k$ and $k+1 < i \le n$, let $\vecg_i = \vecf_{i-1} + \vecg_{i-1}$. Using an inductive argument, and by the construction of $T$, it follows that
\begin{align*}
    T \vecg_i &= T (\vecf_{i-1} + \vecg_{i-1})\\
    &= T \vecf_{i-1} + T \vecg_{i-1}\\
    &= (\vece_i - \vece_{i-1}) + \vece_{i-1}\\
    &= \vece_i.
\end{align*}
It is easy to check that $T T^\top = 4 I$. Finally, we bound the smoothing parameter of the lattice $\Lambda = \ker(T)$. Since $T \in \Z^{n \times (2n + 4)}$ and $T$ has full rank, its kernel $\Lambda$ has dimension $n+4$. The columns of the following matrix give a basis for the lattice $\Lambda$.
\begin{align*}
V =
\begin{bmatrix}
\tilde{Y} &\vece_1 & &-\vece_{k-1} &\\
-\tilde{X} &-\vece_1 & &-\vece_{k-1} &\\
 &1 &1 & &\\
 &1 &-1 & &\\
-\tilde{Z}_{k-1} & & &1 &1\\
-\tilde{Z}_{k-1} & & &1 &-1
\end{bmatrix} \in \Z^{(2n+4)\times (n+4)},
\end{align*}
where we define
\begin{align*}
\tilde{X} &= 
\begin{bmatrix}
-1 & & &\\
1 &-1 & &\\
&\ddots &\ddots &\\
& &1 &-1
\end{bmatrix} \in \Z^{n \times n},\\
\tilde{Y} &= 
\begin{bmatrix}
1 & & &\\
1 &1 & &\\
&\ddots &\ddots &\\
& &1 &1
\end{bmatrix} \in \Z^{n \times n}, \text{ and}\\
\tilde{Z}_{k-1} &= 
\begin{bmatrix}
0 &\ldots &0 & 1 &0 &\ldots &0
\end{bmatrix} \in \Z^{1 \times n}.
\end{align*}
Here $\tilde{Z}_{k-1}$ is the zero matrix except for the $(k-1)$th column which has a $1$ entry. By direct computation, it is easy to see that the columns of $V$ lie in $\ker(T)$. To see that $V$ is a basis for $\ker(T)$, we can show that its columns are linearly independent by constructing a matrix $W \in \Z^{(n+4) \times (2n+4)}$ such that $WV = 2I_{(n+4)\times(n+4)}$. Indeed, we can do so in the following way. We can first define matrices
\[
I_+ = \begin{bmatrix}
1 &         &   &   &   &   &\\
  & \ddots  &   &   &   &   &\\
  &         & 1 &   &   &   &\\
  &         &   & 0 & 1 &   &\\
  &         &   &   & 1 &   &\\
  &         &   &   &   & \ddots &\\
  &         &   &   &   &  & 1\\
\end{bmatrix} \in \Z^{n \times n},\ \ \ 
I_- = \begin{bmatrix}
1 &         &   &   &   &   &\\
  & \ddots  &   &   &   &   &\\
  &         & 1 &   &   &   &\\
  &         &   & 0 & -1 &   &\\
  &         &   &   & 1 &   &\\
  &         &   &   &   & \ddots &\\
  &         &   &   &   &  & 1\\
\end{bmatrix} \in \Z^{n \times n},\]
where the abnormal row is the $(k-1)$th row, and then define
\[ W = \begin{bmatrix}
I_+           & I_-            &   &    &   &\\
              &                & 1 & 1  &   &\\
              &                & 1 & -1 &   &\\
\tilde{Z}_{k} & -\tilde{Z}_{k} &   &    & 1 & 1\\
              &                &   &    & 1 & -1\\
\end{bmatrix} \in \Z^{(n+4) \times (2n+4)},\]
where similarly to before, $\tilde{Z}_k \in \Z^{1 \times n}$ is the one-hot vector with a $1$ in the $k$th column. It is straightforward to verify that $WV = 2I_{(n+4) \times (n+4)}$, showing that the columns of $V$ are linearly independent.

By looking at the columns of $V$, we have $\lambda_{n+4}(\Lambda) \le \sqrt{6}$, so by Lemma~\ref{lemma:smoothing-parameter-estimate}, we have $\eta_\epsilon(\Lambda) \le \sqrt{6} \cdot \sqrt{\omega(\log \secp) + \ln n + \ln t} \le \sigma$, where we set $\epsilon = \negl(\secp)/t$.
Therefore by Lemma~\ref{lemma:mic18-gaussian-projections}, we get that $Q_{]1[}(D_{\Z^{2n+4}, \sigma})$ and $D_{\Z^n, 2\sigma}$ are $\negl(\secp)/t$ close if $\sigma \ge \sqrt{6} \cdot \sqrt{\omega(\log \secp) + \ln n + \ln t}$.
\end{proof}

\begin{lemma}\label{lemma:gadget-Z}
There is a $\poly(n)$ time algorithm that on input $\vecz \in \S_{n,k}$ outputs a matrix $Z \in \Z^{n \times n}$ (as a function of $\vecz$) that satisfies the following properties:
\begin{itemize}
    \item $Z$ is a permutation matrix with signs, i.e. a permutation matrix where the non-zero entries could be $\pm 1$ instead of just $1$,
    \item $Z = Z^\top = Z^{-1}$, and
    \item $Z \vecz = \vecu$.
\end{itemize}
\end{lemma}

\begin{proof}
We can define $Z$ as follows. Let
\begin{align*}
T_{\leq k} &= \{i \in [k] : z_i \neq 0\}, & T_{> k} = \{i \in [n] \setminus [k] : z_i \neq 0\},
\\T^*_{\leq k} &= \{i \in [k] : z_i = 0\}, & T^*_{> k} = \{i \in [n] \setminus [k] : z_i = 0\}.
\end{align*}
Intuitively, $T_{\leq k}$ and $T_{> k}$ partition the non-zero coordinates of $\vecz$ based on whether they lie in the first $k$ coordinates, and $T^*_{\leq k}$ and $T^*_{>k}$ partition the zero-coordinates of $\vecz$ based on whether they lie in the first $k$ coordinates.  Note that by $k$-sparsity of $\vecz$, we have
\[ |T_{>k}| = k - |T_{\leq k}| = |[k] \setminus T_{\leq k}| = |T^*_{\leq k}|.\]
Therefore, we can choose an arbitrary bijection $f : T_{>k} \to T^*_{\leq k}$.

For all $i \in T_{\leq k}$, we set $Z_{i,i} = z_i \in \{+1, -1\}$. For all $i \in T_{>k}^*$, we set $Z_{i,i} = 1$. For all $i \in T_{>k}$, we set $Z_{f(i), i} = z_i \in \{+1, -1\}$ and $Z_{i, f(i)} = Z_{f^{-1}(f(i)), f(i)} = z_i \in \{+1, -1\}$. We set all other entries of $Z$ to be $0$. It's clear from this definition that $Z = Z^\top$.

First, observe that $Z$ is a signed permutation matrix. For all $i \in T_{\leq k} \cup T^*_{>k}$, $Z$ is the identity map up to signs (on basis vectors $\vece_i$), and for all $i \in T_{>k}$, $Z$ consists of signed transpositions $Z \vece_i = z_i \vece_{f(i)}$ and $Z \vece_{f(i)} = z_i \vece_{f^{-1}(f(i))} = z_i \vece_i$. Therefore, $Z$ is a signed permutation matrix, and furthermore we have also shown $Z^2 = I_{n \times n}$. Therefore, $Z = Z^{-1}$.

Lastly, we show $Z \vecz = \vecu$. We can decompose $\vecz$ as $\vecz = \vecz_{\leq k} + \vecz_{>k}$ in the natural way by considering the non-zero coordinates of $\vecz$ on $[k]$ and $[n] \setminus [k]$ respectively. We then have
\[ Z \vecz = Z(\vecz_{\leq k} + \vecz_{>k}) = Z \vecz_{\leq k} + Z \vecz_{>k} = 1_{T_{\leq k}} + 1_{T^*_{\leq k}} = \vecu, \]
as desired.
\end{proof}

\begin{definition}\label{def:mapping}
We define a randomized mapping $\varphi$ as follows. Let $Q$ be as defined in Lemma \ref{lemma:gadget-Q}. We sample $\vecz \sim \S_{n,k}$, $\vecs \sim \Z_q^m$, $\veca \sim \Z_q^{n-1}$, $\vece \sim D_{\Z^m, 2 \sigma}$, $G \sim D_{\Z^{m \times (n+5)}, \sigma}$. Let $Z \in \Z^{n \times n}$ be as defined in Lemma \ref{lemma:gadget-Z} as a function of $\vecz$. On input $B \in \Z_q^{m \times (n-1)}$, we define
\[ \varphi(B; \vecz, \vecs, \veca, \vece, G) = \left[ \left[\vecs, \vecs \cdot \veca^\top + B, G \right]Q^{\top} Z, \vecs + \vece \right]. \]
\end{definition}

First, we show that $\varphi$ maps $B \sim U(\Z_q^{m \times (n-1)})$ to $\LWE(m, \Z_q^n, \S_{n,k}, D_{\Z, \sigma'})$.

\begin{lemma}\label{lemma:phi-maps-uniform-to-k-sparse-LWE}
Assume the same hypothesis as Theorem \ref{thm:regular-lwe-to-k-sparse-ours}. For $B \sim U(\Z_q^{m \times (n-1)})$, we have $\varphi(B)$ and $\LWE(m, \Z_q^n, \S_{n,k}, D_{\Z, \sigma'})$ are $\negl(\secp)$-close.
\end{lemma}

\begin{proof}
We fix $\veca \in \Z_q^{n-1}, \vecz \in S_{n,k}$ and we argue that $\varphi(B)$ maps to $\LWE(m, \Z_q^n, \vecz, D_{\Z, \sigma'})$, i.e. the LWE distribution with secret $\vecz$. Averaging over $\veca$ and $\vecz$ gives the desired result.

First, we show that $X = \left[ \left[\vecs, \vecs \cdot \veca^\top + B, G \right]Q^{\top} Z \right]$ looks uniform. By construction, $[\vecs, \vecs \cdot \veca^\top + B]$ has distribution $U(\Z_q^{m \times n})$, by using the independent randomness of $\vecs$ and $B$. We can write
\[ X = [\vecs, \vecs \cdot \veca^\top + B]  Q_{[n]}^\top Z + G Q_{]n[}^\top Z. \]
Since $Q_{[n]}$ and $Z$ are invertible, by a one-time pad argument, we have $X \sim U(\Z_q^{m \times n})$, independent of $G$ and $e$.

Now, we have to argue that the conditional distribution on $\vecx = \vecs + \vece$ is equal to $X \vecz + \vece'$ for some Gaussian noise $\vece'$. We can directly write
\begin{align*}
\vecx - X \vecz &= \vecs + \vece - ([\vecs, \vecs \cdot \veca^\top + B]  Q_{[n]}^\top Z + G Q_{]n[}^\top Z) \vecz 
\\&= \vecs + \vece - [\vecs, \vecs \cdot \veca^\top + B]  Q_{[n]}^\top \vecu  - G Q_{]n[}^\top \vecu
\\&= \vecs + \vece - [\vecs, \vecs \cdot \veca^\top + B] \vece_1  - G \vecv
\\&= \vece - G \vecv,
\end{align*}
where we use the fact that $Z \vecz = \vecu$, $\vecu^\top Q_{[n]} = \vece_1^\top$ and $\vecu^\top Q_{]n[} = \vecv^\top$.

For all $j \in [m]$, let $\vecg_j \in \Z^{n+5}$ be the $j$th row of $G$. For each entry (row) $\tilde{e}_j$ of $\vece - G \vecv$, we can write $\tilde{e}_j = e_j - \vecg_j^\top \vecv = \inner{[e_j, \vecg_j], [1, -\vecv]}$ and apply Lemma~\ref{lemma:discrete-gaussian-convolution} with the vector $\vecv' = [1, -\vecv]$ to argue that $\tilde{e}_j$ is $O(\epsilon)$-close to $D_{\Z, \sigma'}$ with $\sigma' = \sqrt{(2\sigma)^2 + \sum_{i \in [n+5]} (\sigma v_i)^2} = \sigma \sqrt{4 + \norm{\vecv}_2^2}= 2 \sigma \sqrt{k+1}$, as long as $ \sigma \geq \sqrt{2} \norm{\vecv}_{\infty} \eta_{\epsilon/(2(n+6)^2)}(\Z)$. Now, using the triangle inequality over all $m$ rows to get overall statistical distance $\negl(\secp)$, we can set $\epsilon = \negl(\secp)/m$, for which 
\[\sigma \geq \sqrt{2} \cdot 2 \cdot \eta_{\negl(\secp)/(m n^2)}(\Z) \]
is sufficient. By Lemma \ref{lemma:smoothing-parameter-estimate}, this holds as long as $\sigma \geq 4 \sqrt{\ln m + \ln n + \omega(\log \secp)}$, which we are given.

\end{proof}

Next, we show $\varphi$ maps the standard LWE (with matrices as secrets) to standard LWE in slightly different dimensions, very much following the proof of Claim 3.3 of \cite{Mic18}.

\begin{lemma}\label{lemma:phi-maps-LWE-to-LWE}
Assume the same hypothesis as Theorem \ref{thm:regular-lwe-to-k-sparse-ours}. Let $\cD_1$ denote the distribution of $SA + E \pmod{q}$, where $A \sim U(\Z_q^{\ell \times (n-1)})$, $S \sim U(\Z_q^{m \times \ell})$, $E \sim D_{\Z, \sigma}^{m \times (n-1)}$. Let $\cD_2$ denote the distribution of $\hat{S} \hat{A} + \hat{E} \pmod{q}$, where $\hat{A} \sim U(\Z_q^{(\ell+1) \times (n+1)})$, $\hat{S} \sim U(\Z_q^{m \times (\ell + 1)})$, $\hat{E} \sim D_{\Z, 2\sigma}^{m \times (n+1)}$. Then, $\varphi(\cD_1)$ is $\negl(\secp)$-close to $\cD_2$.
\end{lemma}

The proof goes exactly as in Claim 3.3 of \cite{Mic18}. The only differences are in our matrices $Q, Z$, and our distribution of secrets $\vecz \sim \S_{n,k}$. The full differences are as follows.

\begin{itemize}
    \item While our $Z$ is different, since $Z = Z^\top$ is a permutation matrix with signs, it still holds that $Z \cdot D_{\Z, 2\sigma}^n = D_{\Z, 2\sigma}^n$ due to symmetry.
    \item We have $Q_{]1[}(D_{\Z, \sigma}^{2n+4})$ is $\negl(\secp)/m$-close to $D_{\Z, 2\sigma}^{n}$ by Lemma \ref{lemma:gadget-Q}.
    \item The probability that $\vecw$ (in their notation) is not primitive is at most $\log(q) / 2^\ell = \negl(\secp)$, as desired.
    \item When applying leftover hash lemma (Lemma \ref{lemma:leftover-hash-lemma}), the min-entropy of $\vecz \sim \S_{n,k}$ is now at least $k \log_2(n/k)$. Thus, we require $ k \log_2(n/k) \geq (\ell + 1) \log_2(q) + \omega(\log \secp)$ instead of $n \geq (\ell+1) \log_2(q) + \omega(\log m)$.
\end{itemize}
For completeness, we provide a self-contained proof, exactly following Claim 3.3 of \cite{Mic18}.
\begin{proof}[Proof of Lemma \ref{lemma:phi-maps-LWE-to-LWE}]
Let $B \sim \cD_1$. Let $Y = [\vecs, \vecs \veca^\top + B]$. By linearity, we can decompose $Y$ as $Y = Y_s + Y_e$, where $Y_s = [\vecs, \vecs \veca^\top + S A]$ and $Y_e = [\mathbf{0}, E]$. Similarly, we can write
\[\varphi(B) =  \left[ \left[\vecs, \vecs \cdot \veca^\top + B, G \right]Q^{\top} Z, \vecs + \vece \right] = [X_s, \vecs] + [X_e, \vece],\]
where $X_s = Y_s Q_{[n]}^\top Z$ and $X_e = [Y_e, G] Q^\top Z = [E, G] Q_{]1[}^\top Z$. Our goal is to now show that $[X_s, \vecs]$ is statistically close to $\hat{S} \hat{A}$, and that $[X_e,\vece]$ is statistically close to $\hat{E}$, where $\hat{S} \hat{A} + \hat{E}$ is a sample from $\cD_2$. If this holds, then $\varphi(B)$ is statistically close to $\hat{S} \hat{A} + \hat{E}$, which completes the proof.

First, let us look at $[X_e, \vece]$. Note that $\vece$ is a discrete Gaussian vector of width $2 \sigma$ independent of everything else, so the last column has the desired distribution. Furthermore, note that $E$ and $G$ have entries that are discrete Gaussian of width $\sigma$, so $[E, G] \sim D_{\Z, \sigma}^{m \times (2n+4)}$. By Lemma \ref{lemma:gadget-Q}, setting $t = m$, we can use the triangle inequality over all $m$ rows to get that $[E, G] Q_{]1[}^\top $ is $\negl(\secp)$ close to $D_{\Z, 2 \sigma}^{m \times n}$ as long as $\sigma \geq \sqrt{6} \sqrt{\omega(\log \secp) + \ln n + \ln m}$. Since $Z$ is a signed permutation, by symmetry, we then know that $X_e = [E, G] Q_{]1[}^\top Z$ is $\negl(\secp)$ close to $D_{\Z, 2\sigma}^{m \times n}$, and thus $[X_e, \vece]$ is $\negl(\secp)$ close to $D_{\Z, 2\sigma}^{m \times (n+1)}$, which is the same distribution as $\hat{E}$. Note that this depends only on $\vece, G,$ and $E$.

To finish, we look at $[X_s, \vecs]$. We now define
\[ \hat{S} =  \begin{bmatrix} \vecs, S\end{bmatrix} W^{-1} \in \Z_q^{m \times (\ell + 1)} ,\]
where $W$ is a uniformly random invertible matrix over $\Z_q^{(\ell + 1) \times (\ell + 1)}$. Since $W$ is invertible, using the randomness of $S$ and $\vecs$, $\hat{S}$ is uniformly random independently of $W$. Next, we define
\begin{align*}
\hat{A} &= W H Q_{[n]}^\top Z^\top [ I_{n \times n}, \vecz] \in \Z_q^{(\ell + 1) \times (n + 1)}, \text{ where}
\\H &= \begin{bmatrix}
1 & \veca^\top \\
\mathbf{0} & A
\end{bmatrix} \in \Z_q^{(\ell + 1) \times n}.
\end{align*}
Note that we have the identity $Q^\top_{[n]} Z^\top \vecz  =  Q^\top_{[n]} Z \vecz =Q^\top_{[n]} \vecu = \vece_1$ by Lemmas \ref{lemma:gadget-Z} and \ref{lemma:gadget-Q}, as well as the identity $\hat{S} W H = [\vecs, S] H = Y_s$. Therefore,
\[ \hat{S} \hat{A} = \hat{S} W H  Q_{[n]}^\top Z^\top [ I_{n \times n}, \vecz] = Y_s Q_{[n]}^\top Z^\top [ I_{n \times n}, \vecz] = [Y_s Q_{[n]}^\top Z, Y_s \vece_1] = [X_s, \vecs], \]
as desired.

Now, we have to show that $\hat{S}$ and $\hat{A}$ have the correct distributions. We have already shown that $\hat{S}$ has the correct distribution (only depending on $S$ and $\vecs$), so it suffices to show that $\hat{A}$ has the correct distribution given $S$ and $\vecs$, using the randomness of $A, \veca, W$ and $\vecz$. First, let's look at the matrix $WH$. Let $\vecw$ be the first column of $W$. The first column of $WH$ will be exactly $\vecw$. Since $W$ is a uniformly random invertible matrix, $\vecw$ is distributed uniformly among all primitive vectors in $\Z_q^{\ell+1}$, i.e. so that $\gcd(\vecw, q) = 1$. By Lemma \ref{lemma:primitive-vector-negl}, as long as $\log(q)/2^\ell = \negl(\secp)$, which we have assumed, then the distribution of $\vecw$ is $\negl(\secp)$-close to uniform over $\Z_q^{\ell + 1}$. The remaining columns of $WH$ will be $W \begin{bmatrix} \veca^\top \\ A \end{bmatrix}$, which by using the uniform randomness of $\veca$ and $A$, and the invertibility of $W$, will be uniformly random and independent of $\vecw$. Therefore, $WH \in \Z_q^{(\ell + 1) \times n}$ is $\negl(\secp)$-close to uniformly random. Now, since $Q_{[n]}^\top$ and $Z^\top$ are invertible, we have $WHQ^\top_{[n]} Z^\top$ is $\negl(\secp)$-close to uniform, independently of $\vecz$. Let $A' = W H Q_{[n]}^\top Z^\top$, which we have just shown is $\negl(\secp)$-close to uniform, independently of $\vecz$. Note that
\[ \hat{A} = A'[ I_{n \times n}, \vecz] = [A', A'\vecz ].\]
Applying the leftover hash lemma (Lemma \ref{lemma:leftover-hash-lemma}) and Lemma \ref{lemma:min-entropy-of-sparse-secrets}, since $k \log_2(n/k) \geq (\ell + 1) \log_2(q) + \omega(\log \secp)$, we know $\hat{A}$ is $\negl(\secp)$-close to uniform, independently of $\hat{S}$ and $\hat{E}$. This completes the proof that $\varphi(\cD_1)$ and $\cD_2$ are $\negl(\secp)$-close. 
\end{proof}

With the above claims, we are ready to prove the main theorem of this section.

\begin{proof}[Proof of Theorem \ref{thm:regular-lwe-to-k-sparse-ours}]
We will show the contrapositive. Suppose we have a $T$-time distinguisher between $\LWE(m, \Z_q^n, \S_{n,k}, D_{\Z, \sigma'})$ and $U(\Z_q^{m \times n} \times \Z_q^m) = U(\Z_q^{m \times (n+1)})$ with advantage $2 \epsilon$.

We have two cases. Suppose that this distinguisher distinguishes between $U(\Z_q^{m \times n} \times \Z_q^m) = U(\Z_q^{m \times (n+1)})$ and $\cD_2$ as given in Lemma \ref{lemma:phi-maps-LWE-to-LWE}, with advantage $\epsilon$. Then, we have a $T$ time distinguisher between $\LWE(n+1, \Z_q^{\ell + 1}, \Z_q^{m \times (\ell + 1)}, D_{\Z^m, 2\sigma})$ and $U(\Z_q^{(\ell +1) \times (n+1)} \times \Z_q^{m \times (n+1)})$ where we simply discard the samples, i.e. the first part in $\Z_q^{(\ell + 1) \times (n+1)}$ (the matrix $\hat{A}$).

Now, for the second case, suppose that this distinguisher does not distinguish between $U(\Z_q^{m \times n} \times \Z_q^m) = U(\Z_q^{m \times (n+1)})$ and $\cD_2$ with advantage $\epsilon$. Then, we have a $T$-time distinguisher between $\LWE(m, \Z_q^n, \S_{n,k}, D_{\Z, \sigma'})$ and $\cD_2$ with advantage $\geq 2 \epsilon - \epsilon = \epsilon$ by the triangle inequality. Now, we can use this distinguisher to distinguish $\LWE(n-1, \Z_q^{\ell}, \Z_q^{m \times \ell}, D_{\Z^m, 2\sigma})$ and $U(\Z_q^{\ell \times (n-1)} \times \Z_q^{m \times (n-1)})$ by once again discarding the samples, i.e. the first part in $\Z_q^{\ell \times (n-1)}$ (the matrix $A$), and then by applying $\varphi$ to the remaining part in $\Z_q^{m \times (n-1)}$. Now, using Lemmas \ref{lemma:phi-maps-uniform-to-k-sparse-LWE} and \ref{lemma:phi-maps-LWE-to-LWE}, the resulting distributions coming out of $\varphi$ when given $U(\Z_q^{m \times (n-1)})$ and $\cD_1$ will be $\negl(\secp)$-close to $\LWE(m, \Z_q^n, \S_{n,k}, D_{\Z, \sigma'})$ and $\cD_2$, respectively. Thus, our assumed distingiusher will be correct, where the only runtime increase is in the randomized transformation $\varphi$, taking time $\poly(n,m, \log(q), \log(\secp))$.
\end{proof}

Now, we state a simpler version of Theorem \ref{thm:regular-lwe-to-k-sparse-ours} that is easier to use.

\begin{corollary}\label{cor:useful-lwe-to-k-sparse-lwe}
Suppose $\log(q)/2^\ell = \negl(\secp), \sigma \geq 4 \sqrt{\omega(\log \secp) + \ln n + \ln m}$, and $k \log_2(n/k) \geq (\ell + 1) \log_2(q) + \omega(\log \secp)$. Then, if $\LWE(n, \Z_q^\ell, \Z_q^\ell, D_{\Z, \sigma})$ and $U(\Z_q^{\ell \times n} \times \Z_q^{n})$ have no $T + \poly(n, m, q, \secp)$ time distinguisher with advantage $\epsilon$, then $\LWE(m, \Z_q^n, \S_{n,k}, D_{\Z, \sigma'})$ and $U(\Z_q^{n \times m} \times \Z_q^m)$ have no $T$-time distinguisher with advantage $2\epsilon m + \negl(\secp)$, where $\sigma' = 2 \sigma \sqrt{k+1}$.
\end{corollary}

\begin{proof}
If $\LWE(n, \Z_q^\ell, \Z_q^\ell, D_{\Z, \sigma})$ and $U(\Z_q^{\ell \times n} \times \Z_q^{n})$ cannot be distinguished with advantage $\epsilon$, then by a hybriding argument, the version where the secrets are matrices (with dimension $m$ instead of $1$) cannot be distinguished with advantage $\epsilon m$. Then, applying Theorem \ref{thm:regular-lwe-to-k-sparse-ours}, $\LWE(m, \Z_q^n, \S_{n,k}, D_{\Z, \sigma'})$ and $U(\Z_q^{n \times m} \times \Z_q^m)$ cannot be distinguished with advantage $2\epsilon m + \negl(\secp)$, where we reparameterize to absorb small additive factors, with the observation that $\LWE$ is harder when the dimension and noise grow, and easier when the number of samples grows.
\end{proof}

%% file: LWEtoCLWE.tex
\section{Reducing LWE to CLWE}\label{sec:mainreduction}

Our main result in this section is a reduction from decisional fixed-norm LWE to decisional CLWE:

\begin{theorem}[Fixed-Norm LWE to CLWE]\label{thm:fixed-norm-lwe-clwe}
Let $r \in \R_{\geq 1}$, and let $\S$ be an arbitrary distribution over $\Z^n$ where all elements in the support of $\S$ have $\ell_2$ norm $r$. Then, for 
\begin{align*}
\gamma &= r \cdot \sqrt{\ln(m) + \ln(n) + \omega(\log \secp)}, \text{  and}
\\\beta &= O \left( \frac{\sigma}{q} \right),
\end{align*}
if there is no $T + \poly(n, m, \log(q), \log(\sigma), \log(\secp))$ time distinguisher between $\LWE(m, \Z_q^n, \S, D_{\Z, \sigma})$ and $U(\Z_q^{n \times m} \times \Z_q^m)$ with advantage at least $\epsilon - \negl(\secp)$, then there is no $T$-time distinguisher between $\CLWE(m, D_1^n, \frac{1}{r} \cdot \S, \gamma, \beta)$ and $D_1^{n \times m} \times U(\T^m)$ with advantage $\epsilon$, as long as $\sigma \geq 3r \sqrt{\ln(m) + \ln(n) + \omega(\log \secp)}$.
\end{theorem}

See Figure~\ref{fig:steps-copy} for a summary of the steps. We note that the dimension and number of samples remains the same in this reduction, and the advantage stays the same up to additive $\negl(\secp)$ factors. We also remark that to keep the theorem general, the final distribution is not exactly the CLWE distribution, as the secret distribution is $\frac{1}{r} \cdot \S$ instead of $U(S^{n-1})$. However, using Lemma~\ref{lemma:wc-to-ac-clwe}, it is straightforward to reduce from $\frac{1}{r} \cdot \S$ secrets to $U(S^{n-1})$ secrets.

This reduction goes via a series of transformations, which we briefly outline below:

\begin{enumerate}
    \item Starting from standard decisional LWE, with samples $\veca \sim U(\Z_q^n)$, (fixed) secret $\vecs \sim \S$ (where the support of $\S$ has fixed norm), and errors $e \sim D_{\Z, \sigma}$, we convert discrete Gaussian errors $e$ to continuous Gaussian errors $e \sim D_{\sigma_2}$ for $\sigma_2$ slightly larger than $\sigma$. 
    \item We convert discrete uniform samples $\veca \sim U(\Z_q^n)$ to continuous uniform samples $\veca \sim U(\T_{q}^n)$ with errors from $D_{\sigma_3}$, where $\sigma_3$ is slightly larger than $\sigma_2$.
    \item We convert uniform $\veca \sim U(\T_{q}^{n})$ to Gaussian $\veca \sim D_{1}^{n}$; viewing it as a CLWE distribution, we scale such the secret $\vecs$ is a unit vector (i.e. $\vecs \sim \frac{1}{r} \cdot \S$), $\gamma \approx r$, and the noise distribution becomes $D_{\beta}$ where $\beta = \sigma_3/q$.
\end{enumerate}

\paragraph{Setting of parameters.} If we start with dimension $n$ and $m$ samples with error width $\sigma$:
\begin{enumerate}
    \item After the first step, we get $\sigma_2 = O(\sigma)$, as long as $\sigma \geq 2\sqrt{\ln m + \omega(\log \secp)}$.
    \item After the second step, we get $\sigma_3 =O(\sigma_2) = O(\sigma)$, as long as $\sigma_2 \geq 3 r \sqrt{\ln n + \ln m + \omega(\log \secp)}$.
    \item After the third step, we get $\gamma = r \cdot \sqrt{\ln n + \ln m + \omega(\log \secp)}$ and $\beta = \sigma_3/q = O(\sigma/q)$.
\end{enumerate}

\paragraph{Step 1: Converting discrete errors to continuous errors.} First, we make the error distribution statistically close to a continuous Gaussian instead of a discrete Gaussian. Essentially, all we do is add a small continuous Gaussian noise to the second component and argue that this makes the noise look like a continuous Gaussian instead of a discrete one.

This sort of reduction is standard in the literature, but we provide it here for completeness.

\begin{lemma}\label{lemma:discrete-to-continuous-errors}
Let $n, m, q \in \N, \sigma \in \R_{>0}$, and suppose $\sigma > \sqrt{4 \ln m + \omega(\log \secp)}$. For any distribution $\S$ over $\Z^n$, suppose there is no distinguisher between $\LWE(m, \Z_q^n, \S, D_{\Z, \sigma})$ and $U(\Z_q^{n \times m} \times \Z_{q}^m)$ running in time $T + \poly(m,n, \log(q), \log(\sigma))$. Then, there is no $T$-time distinguisher $\LWE(m, \Z_q^n, \S, D_{\sigma'})$ and $U(\Z_q^{n \times m}) \times U(\T_{q}^m)$ with an additive $\negl(\secp)$ advantage loss, where
\[ \sigma' = \sqrt{\sigma^2 + 4\ln(m) + \omega(\log \secp)} = O(\sigma).\]
\end{lemma}

\begin{proof}
We run our original distinguisher for $\LWE(m, \Z_q^n, \S, D_{\sigma'})$ and $U(\Z_q^{n \times m}) \times U(\T_{q}^m)$. For every sample $(\veca, b)$ (from either $\LWE(m, \Z_q^n, \S, D_{\Z, \sigma})$ or $U(\Z_q^{n \times m} \times \Z_{q}^m)$), we sample a continuous Gaussian $e' \sim D_{\sigma''}$ where $\sigma''$ will be set later, and send $(\veca, b + e' \pmod{q})$ to the distinguisher.

By Lemma \ref{lemma:gaussian-mod-lattice-looks-uniform}, we know that the distribution of $e' \pmod{1}$ has statistical distance at most $\epsilon$ to $U([0,1))$ as long as $\sigma'' \geq \eta_{\epsilon}(\Z)$. Therefore, if we are given samples from $U(\Z_q^{n \times m} \times \Z_{q}^m)$, due to symmetry of $b \sim \Z_q$, we can set $\epsilon = \secp^{- \omega(1)}/m$ to have $b + e' \pmod{q}$ look $\negl(\secp)/m$-close to $\T_{q}$, making it look like samples from $U(\Z_q^{n \times m}) \times U(\T_{q}^m)$.

If we are given samples from $\LWE(m, \Z_q^n, \S, D_{\Z, \sigma})$, then the second component can be seen as having noise $e + e'$, where $e \sim D_{\Z, \sigma}$ and $e' \sim D_{\sigma''}$. Applying Lemma \ref{lemma:discrete-plus-continuous}, as long as $1/\sqrt{1/\sigma^2 + 1/(\sigma'')^2} \geq \eta_{\epsilon}(\Z)$, then $e + e'$ will look $O(\epsilon)$-close to $D_{\sqrt{\sigma^2 + (\sigma'')^2}}$. Thus, as long as $\sigma, \sigma'' \geq \sqrt{2} \cdot \eta_{\epsilon}(\Z)$, it all goes through, as taking errors mod $q$ (i.e. in $\T_q$ instead of $\R$) can only decrease statistical distance. Now, applying Lemma \ref{lemma:smoothing-parameter-estimate}, we can set $\epsilon = \secp^{- \omega(1)}/m$ and $\sigma'' = \sqrt{4 \ln(m) + \omega(\log \secp)}$, and as long as $\sigma > \sqrt{4 \ln(m) + \omega(\log \secp)}$, all goes through. Now, doing the triangle inequality over all $m$  samples, we get $\negl(\secp)$-closeness of all samples.
\end{proof}

\paragraph{Step 2: Converting discrete to continuous samples.} Now, we convert discrete uniform samples $\veca \sim \Z_{q}^n$ to continuous uniform samples $\veca \sim \T_{q}^n$.
\begin{lemma}\label{lemma:discrete-to-continuous-samples}
Let $n, m, q \in \N$, $\sigma \in \R$. Let $\S$ be a distribution over $\Z^n$ where all elements in the support have fixed norm $r$, and suppose that
\[\sigma \geq 3r \sqrt{\ln n + \ln m + \omega(\log \secp)}. \]
Suppose there is no $T + \poly(m, n, \log(q), \log(\sigma))$-time distinguisher between the distributions $\LWE(m, \Z_q^n, \S, D_{\sigma})$ and $U(\Z_q^{n \times m}) \times U(\T_q^m)$. Then, there is no $T$-time distinguisher between the distributions $\LWE(m, \T_q^n, \S, D_{\sigma'})$ and $U(\T_q^{n \times m} \times \T_q^m)$ with an additive $\negl(\secp)$ advantage loss, where we set
\[ \sigma' = \sqrt{\sigma^2 + 9r^2 (\ln n + \ln m + \omega(\log \secp))} = O(\sigma). \]
\end{lemma}

\begin{proof}
We run our distinguisher for $\LWE(m, \T_q^n, \S, D_{\sigma'})$ and $U(\T_q^{n \times m} \times \T_q^m)$. Let $\epsilon = \negl(\secp)/m$, and let $\sigma'' \geq \sqrt{2} \cdot \eta_{\epsilon}(\Z^n)$. For each sample $(\veca,b)$ (from either $\LWE(m, \Z_q^n, \S, D_{\sigma})$ or $U(\Z_q^{n \times m}) \times U(\T_q^m)$), we sample a continuous Gaussian $\veca' \sim \left(D_{\sigma''}\right)^{n}$ and send $(\veca + \veca' \pmod{q}, b)$ to the distinguisher. By Lemma \ref{lemma:gaussian-mod-lattice-looks-uniform}, we know that the distribution of $\veca' \pmod{1}$ has statistical distance at most $\epsilon = \negl(\secp)/m$ to $U([0,1)^{n})$. Thus, by symmetry over $\veca \sim (\Z_{q})^n$, the distribution of $\veca + \veca' \pmod{q}$ will be $\negl(\secp)/m$-close to uniform over $(\T_{q})^n$. Therefore, by the triangle inequality, if we are given samples from $U(\Z_q^{n \times m}) \times U(\T_q^m)$, the reduction gives samples to the distinguisher that are $\negl(\secp)$-close to $U(\T_q^{n \times m} \times \T_q^m)$.

If we are given samples from $\LWE(m, \Z_q^n, \S, D_{\sigma})$, then the reduction gives us (taking everything mod $q$)
\[  (\veca + \veca', \inner{\veca, \vecs} + e) = (\veca + \veca', \inner{\veca + \veca', \vecs} + e - \inner{\veca', \vecs}) = (\veca + \veca', \inner{\veca + \veca', \vecs} + e'), \]
where we define 
\[e' = e - \inner{\veca', \vecs} \]
over $\R$. Conditioned on $\veca + \veca' \mod q$, $\veca'$ is a discrete Gaussian distributed according to $D_{\Z^n + (\veca + \veca'), \sigma''}$. By Lemma \ref{lemma:discrete-plus-continuous}, as long as $\sigma \ge r \sigma''$, the distribution of $e'$ is $O(\epsilon) = \negl(\secp)/m$ close to $D_{\sigma'}$, where 
\begin{align*}
    \sigma' = \sqrt{\sigma^2 + r^2 (\sigma'')^2}.
\end{align*}
Averaging the distribution of $e'$ over $\vecs$ will not change the distribution over $e'$, as all secrets $\vecs$ have fixed norm $r$. Therefore, if we are given the $m$ samples from $\LWE(m, \Z_q^n, \S, D_{\sigma})$, the reduction gives us samples $\negl(\secp)$-close to $\LWE(m, \T_q^n, \S, D_{\sigma'})$, as desired.

To set parameters, we choose $\sigma'' = 3 \sqrt{\ln n + \ln m + \omega(\log \secp)}$ to ensure that $\sigma'' \geq \sqrt{2} \cdot \eta_{\negl(\secp)/m}(\Z^n)$. This gives
\[ \sigma' = \sqrt{\sigma^2 + 9r^2 (\ln n + \ln m + \omega(\log \secp))}, \]
along with the requirement that
\[\sigma \geq r \sigma'' = 3r \sqrt{\ln n + \ln m + \omega(\log \secp)}. \]
\end{proof}

\paragraph{Step 3: Converting uniform to Gaussian samples.}

\begin{lemma}\label{lemma:gaussian-preimage-sampling}
Let $t \in \R_{>0}$ be a parameter. There is a $\poly(n,\log(t), \log(\secp))$-time algorithm such that on input $\vecz \in \T_{1}^n$, the algorithm outputs some $\vecy \in \R^{n}$ such that $\vecy = \vecz \pmod{1}$. Moreover, if $\vecz \sim U(\T_{1}^n)$, then the distribution on the outputs $\vecy$ is $\negl(\secp)/t$-close to $D_\tau^n$, where $\tau = \sqrt{\ln n + \ln t + \omega(\log \secp)}$.
\end{lemma}

\begin{remark}
In the discrete setting, there is in some sense a necessary multiplicative $\Omega(\log q)$ overhead in the dimension due to entropy arguments, but the above shows that we can overcome that barrier in the continuous case.
\end{remark}

\begin{proof}
We give each coordinate of $\vecy$ separately. By the triangle inequality, it suffices to show how to sample $y \in \R$ such that $y = z \pmod{1}$ and such that if $z \sim \T_{1}$, then $y$ is $\negl(\secp)/(tn)$-close to $D_{\tau}$. We sample
\[ y \sim D_{\Z + z, \tau}, \]
which can be sampled efficiently (see e.g. \cite{brakerski2013classical}, Section~5.1 of full version), where we have $\negl(\secp)/(tn)$ statistical distance between $y$ and $D_{\Z + z, \tau}$, and always satisfy $y \in \Z + z$. Since $y \in \Z + z$, it follows that $y = z \pmod{1}$.

Now, we need to argue that the distribution of $y$ looks $\negl(\secp)/(tn)$-close to $D_{\tau}$ when $z \sim U(\T_1)$.  Note that for fixed $z \in [0,1)$, we can write the generalized PDF of $D_{\Z+z, \tau}$ as
\[ D_{\Z + z, \tau}(x) = \delta(x - z \mod{1}) \cdot \frac{\rho_{\tau}(x)}{\rho_{\tau}(\Z + z)} \]
for arbitrary $x \in \R$, where $\delta(\cdot)$ is the Dirac delta function. Thus, as long as $\tau \geq \eta_{\epsilon}(\Z)$ (for $\epsilon$ set later), the density of the marginal distribution $D_{\Z + z, \tau}$ where $z \sim U([0,1))$ is given by
\begin{align*} D_{\Z + U([0,1)), \tau}(x) &= \int_0^1 1 \cdot D_{\Z + z, \tau}(x) \cdot dz
\\&= \int_0^1 \delta(x - z \mod{1}) \cdot \frac{\rho_{\tau}(x)}{\rho_{\tau}(\Z + z)} dz
\\&= \frac{\rho_{\tau}(x)}{\rho_{\tau}(\Z + x)}
\\&\in \left[1, \frac{1 + \epsilon}{1 - \epsilon} \right] \cdot \frac{\rho_{\tau}(x)}{ \rho_{\tau}(\Z)}
\\&\propto \left[1, \frac{1 + \epsilon}{1 - \epsilon} \right] \cdot \rho_{\tau}(x),
\end{align*}
where the inclusion comes from Lemma \ref{lemma:smoothing-swallows-shifting}.
Therefore, a standard calculation shows that the statistical distance between $D_{\Z + U([0,1)), \tau}$ and $D_\tau$ is at most $O(\epsilon)$. Setting $\epsilon = \secp^{-\omega(1)}/(t \cdot n)$, we need to take $\tau \geq \eta_{\secp^{-\omega(1)}/(t \cdot n)}(\Z)$, which we can do by setting $\tau = \sqrt{\ln n + \ln t + \omega(\log \secp)}$ by Lemma \ref{lemma:smoothing-parameter-estimate}.

\end{proof}

\begin{lemma}\label{lemma:uniform-to-gaussian-samples}
Let $n,m,q \in \N, \sigma, r, \gamma \in \R$. Let $\S$ be a distribution over $\Z^n$ where all elements in the support have fixed norm $r$. Suppose there is no $T + \poly(n,m, \log(q), \log(\secp))$ time distinguisher between the distributions $\LWE(m, \T_q^n, \S, D_\sigma)$ and $U(\T_q^{n \times m} \times \T_q^m)$. Then, there is no $T$-time distinguisher between the distributions $\CLWE(m, D_1^n, \frac{1}{r} \cdot \S, \gamma, \beta)$ and $D_1^{n \times m} \times U(\T_1^m)$ with an additive advantage loss of $\negl(\secp)$, where
\begin{align*}
    \gamma &= r \cdot \sqrt{\ln n + \ln m + \omega(\log \secp)},
    \\\beta &= \frac{\sigma}{q}.
\end{align*}
\end{lemma}

\begin{proof}
We run the distinguisher for $\CLWE(m, D_1^n, \frac{1}{r} \cdot \S, \gamma, \beta)$ and $D_1^{n \times m} \times U(\T_1^m)$. For each sample $(\veca,b)$ from either $\LWE(m, \T_q^n, \S, D_\sigma)$ or $U(\T_q^{n \times m} \times \T_q^m)$, we invoke Lemma~\ref{lemma:gaussian-preimage-sampling} on $\veca/q$ with parameter $t = m$ to get some $\vecy \in \R^n$ with statistical distance $\negl(\secp)/m$ from $D_\tau^n$ such that $\vecy = \veca/q \pmod{1}$, where $\tau = \sqrt{\ln n + \ln m + \omega(\log \secp)}$. We then send $(\vecy/\tau, b/q)$ to the distinguisher. Let $\gamma = r \cdot \tau$, $\vecy' = \vecy/\tau$, $\vecs' = \vecs/r$, and $e' = e/q$. If $(\veca, b)$ is a sample from $\LWE(m, \T_q^n, \S, D_{\sigma})$, then for secret $\vecs \sim \S$, since $\vecs \in \Z^n$, we have
\begin{align*}
(\vecy/\tau , b/q) = (\vecy', \inner{\veca/q, \vecs} + e/q \pmod{1}) &= (\vecy', \inner{\vecy, \vecs} + e' \pmod{1}) 
\\&= (\vecy', r \cdot \tau \cdot \inner{ \vecy', \vecs/r} + e' \pmod{1})
\\&= (\vecy', \gamma \cdot \inner{ \vecy', \vecs'} + e' \pmod{1})
\end{align*} 
where this is now $\negl(\secp)/m$ close to a sample from  $\CLWE(m, D_1^n, \frac{1}{r} \cdot \S, \gamma, \beta)$, as $\vecy' \sim D_1^n$, $\vecs' \sim \frac{1}{r} \cdot \S$, and $e' \sim D_{\sigma/q} = D_{\beta}$. Applying this reduction to $U(\T_q^{n \times m} \times \T_q^m)$ clearly gives us a statistically close sample to $D_1^{n \times m} \times U(\T_1^m)$ by Lemma \ref{lemma:gaussian-preimage-sampling} and the triangle inequality over all $m$ samples.
\end{proof}

\paragraph{Step 4 (optional): Converting the secret to a random direction.} The distribution on the secret as given above is not uniform over the sphere, so if desired, one can apply the worst-case to average-case reduction for CLWE (\cite{CLWE}, Claim 2.22). For completeness, we provide a proof.

\begin{lemma}[\cite{CLWE}, Claim 2.22]\label{lemma:wc-to-ac-clwe}
Let $n,m \in \N$, and let $\beta \in \R_{>0}$. Let $\S$ be a distribution over $\R^n$ of fixed norm $1$. There is no $T$-time distinguisher between the distributions $\CLWE(m, D_1^n, \gamma, \beta)$ and $D_{1}^{n \times m} \times U(\T_1^m)$, assuming there is no $T + \poly(n,m)$ time distinguisher between the distributions $\CLWE(m, D_1^n, \S, \gamma, \beta)$ and $D_{1}^{n \times m} \times U(\T_1^m)$. 
That is, we can reduce CLWE to CLWE to randomize the secret to be a uniformly random unit vector instead of drawn from (possibly discrete) $\S$.
\end{lemma}

Note that while we do not use Lemma~\ref{lemma:wc-to-ac-clwe} in proving Theorem~\ref{thm:fixed-norm-lwe-clwe}, we do use the lemma in subsequent sections.

\begin{proof}
We run the distinguisher for $\CLWE(m, D_1^n, \gamma, \beta)$ and $D_{1}^{n \times m} \times U(\T_1^m)$. Let $R \in \R^{n \times n}$ be a uniformly random rotation matrix in $\R^n$, fixed for all samples. When giving the distinguisher a sample, we get $(\veca, b)$ from either $\CLWE(m, D_1^n, \S, \gamma, \beta)$ or $D_{1}^{n \times m} \times U(\T_1^m)$, and send $(R \veca, b)$ to the distinguisher. If $(\veca, b)$ is drawn from $\CLWE(m, D_1^n, \S, \gamma, \beta)$, then we have
\begin{align*} (R\veca, b) = (R \veca, \gamma \inner{\veca, \vecs} + e \pmod{1}) &= (R \veca, \gamma \inner{R\veca, R\vecs} + e \pmod{1}) 
\\&= (\veca', \gamma \inner{\veca', \vecw} + e \pmod{1}),
\end{align*}
for $\veca \sim D_{1}^n$, $\vecs \sim \S$, and $e \sim D_{\beta}$,
where we set $\veca' = R \veca$ and $\vecw = R \vecs$ (fixed for all samples). For an arbitrary rotation $R$, since the distribution on $\veca$ is spherically symmetric, we have $\veca' = R \veca \sim D_1^n$, independently of $R$. For a random rotation matrix $R$, for arbitrary $\vecs$, we have that $\vecw = R \vecs$ is a uniformly random unit vector in $\R^n$. Since this holds for arbitrary $\vecs$, this also holds when averaging over the distribution $\vecs \sim \S$. If $(\veca, b)$ is drawn from $D_{1}^{n \times m} \times U(\T_1^m)$, then $(R \veca, b)$ is drawn identically to $(\veca, b)$, since the distribution on $\veca' = R \veca$ is spherically symmetric. Thus, the reduction maps the distributions perfectly.
\end{proof}

Now, we are ready to prove the main theorem of this section, Theorem~\ref{thm:fixed-norm-lwe-clwe}.

\begin{proof}[Proof of Theorem~\ref{thm:fixed-norm-lwe-clwe}]
Throughout this proof, when we refer to distinguishing probability, we omit additive $\negl(\secp)$ terms for simplicity.

Suppose there is no distinguisher with advantage $\epsilon$ between $\LWE(m, \Z_q^n, \S, D_{\Z, \sigma})$ and $U(\Z_q^{n \times m} \times \Z_q^m)$. Then, by Lemma~\ref{lemma:discrete-to-continuous-errors}, there is no $\epsilon$-distinguisher between $\LWE(m, \Z_q^n, \S, D_{\sigma_2})$ and $U(\Z_q^{n \times m}) \times U(\T_q^m)$, where $\sigma_2 = O(\sigma)$, as long as $\sigma \geq \sqrt{4\ln(m) + \omega(\log \secp)}$, which it is by our assumption on $\sigma$. Then, by Lemma~\ref{lemma:discrete-to-continuous-samples}, there is no $\epsilon$-distinguisher between $\LWE(m, \T_q^n, \S, D_{\sigma_3})$ and $U(\T_q^{n \times m} \times \T_q^m)$, where $\sigma_3 = O(\sigma_2) = O(\sigma)$, which holds as long as $\sigma_2 \geq 3r \sqrt{\ln(m) + \ln(n) + \omega(\log \secp)}$, which it does because $\sigma_2 \geq \sigma \geq 3r \sqrt{\ln(m) + \ln(n) + \omega(\log \secp)}$.
Now, by Lemma~\ref{lemma:uniform-to-gaussian-samples}, there is no $\epsilon$-distinguisher between $\CLWE(m, D_1^n, \frac{1}{r} \cdot \S, \gamma, \beta)$ and $D_{1}^{n \times m} \times U(\T_1^m)$, where
\[ \gamma = r \cdot \sqrt{\ln(m) + \ln(n) + \omega(\log \secp)},\]
and
\[ \beta = \frac{\sigma_3}{q} = O\left( \frac{\sigma}{q} \right), \]
as desired.
\end{proof}

\subsection{Full Reduction from LWE to CLWE}

Now, to reduce from standard decisional LWE where the secret is drawn uniformly over $\Z_q^n$ instead of a fixed-norm distribution, we need to somehow reduce standard LWE to some version where the norm is fixed. We show two ways to do this:
\begin{enumerate}
    \item In Corollary~\ref{cor:lwe-to-clwe}, we use a reduction from LWE to binary-secret LWE \cite{Mic18} (i.e. Section~\ref{sec:k-sparse-lwe-hardness} but without sparsity) to bridge this gap.
    \item In Appendix~\ref{appendix:alternate-lwe-to-clwe}, we give another (perhaps simpler) reduction, but we reduce to search CLWE instead of decisional CLWE. (As a result of Appendix~\ref{appendix:clwe-to-lwe}, we get an indirect search-to-decision reduction for discrete-secret CLWE that can be applied here.)
\end{enumerate}
In this section, we show the first approach.

\begin{theorem}[\cite{Mic18}, Theorem~3.1 and Lemma~2.9]\label{thm:uniform-to-binary-secrets} 
Let $q, \ell, n, m \in \Z$, $\sigma \in \R$. There is no $T$-time algorithm has advantage $\epsilon$ in distinguishing $\LWE(m, \Z_q^{n+1}, \{+1, -1\}^{n+1}, D_{\Z, \sigma'})$ and $U(\Z_q^{(n+1) \times m} \times \Z_q^{m})$, assuming there is no time $T + \poly(\ell, n, \log(q), \log(\secp))$ algorithm with advantage $(\epsilon - \negl(\secp))/(2m)$ in distinguishing $\LWE(n+1, \Z_q^\ell, \Z_q^\ell, D_{\Z, \sigma})$ and $U(\Z_q^{\ell \times (n+1)} \times \Z_q^{n+1})$, as long as $\log(q) / 2^{\ell} = \negl(\secp), \sigma \geq 4\sqrt{\omega(\log \secp) + \ln n + \ln m}$, $n \geq 2\ell \log_2 q + \omega(\log \secp)$, and $\sigma' = 2 \sigma \sqrt{n+1}$.
\end{theorem}

\begin{remark}
Note that we phrase the parameter requirements differently here than is done in \cite{Mic18}, mainly because we want to delink the security parameter from $n$. Explicitly:
\begin{itemize}
    \item The requirements $q \leq 2^{\poly(n)}$ and $\ell \geq \omega(\log n)$ in \cite{Mic18} are needed only to make sure that the first row of a primitive matrix is close to uniform over $\Z_q$. Indeed, Lemma 2.2 of \cite{Mic18} shows the statistical distance is at most $\log(q)/2^\ell$. Thus, the requirement $\log(q) / 2^\ell = \negl(\secp)$ is sufficient.
    \item We require $\sigma \geq 4\sqrt{\omega(\log \secp) + \ln n + \ln m}$ instead of $\sigma \geq \omega(\sqrt{ \log n})$ for various triangle inequalities to go through to get $\negl(\secp)$ overall statistical distance.
\end{itemize}
\end{remark}

Now, we are ready to give a proof of Corollary~\ref{cor:lwe-to-clwe}.
\begin{corollary}[Full Reduction from LWE to CLWE]
\label{cor:lwe-to-clwe}
Let $q,\ell,n,m \in \N$ with $m > n$, and let $\gamma, \beta, \sigma, \epsilon \in \R_{>0}$. There is no $T$-time distinguisher with advantage $\epsilon$ between $\CLWE(m, D_1^{n}, \gamma, \beta)$ and $D_1^{n \times m} \times U(\T^{m})$, assuming there is no $T + \poly(\ell,n, m, \log(q), \log(\sigma), \log(\secp))$ time distinguisher with advantage $(\epsilon - \negl(\secp))/(2m)$ between $\LWE(m, \Z_q^\ell, \Z_q^\ell, D_{\Z, \sigma})$ and $U(\Z_q^{\ell \times m} \times \Z_q^m)$, for
\begin{align*}
\gamma &= O\left(\sqrt{n} \cdot \sqrt{\ln m + \omega(\log \secp)} \right),
\\ \beta &= O \left( \frac{\sigma \sqrt{n}}{q} \right),
\end{align*}
as long as $\log(q)/2^{\ell} = \negl(\secp)$, $n \geq 2\ell \log_2 q + \omega(\log \secp)$, and $\sigma \geq C \cdot \sqrt{ \ln m + \omega(\log \secp)}$ for some universal constant $C$.
\end{corollary}

\begin{remark}
Note that for reasonable parameter settings of CLWE (namely where $\beta \ll 1$), we require $q/\sigma \gg \sqrt{n}$.
\end{remark}

\begin{proof}[Proof of Corollary~\ref{cor:lwe-to-clwe}]
Suppose there is no distinguisher with advantage $(\epsilon - \negl(\secp))/(2m)$ between $\LWE(m, \Z_q^\ell, \Z_q^\ell, D_{\Z, \sigma})$ and $U(\Z_q^{\ell \times m} \times \Z_q^m)$. Then, since $n < m$ and more samples can only help, there is no distinguisher with advantage $(\epsilon - \negl(\secp))/(2m)$ between $\LWE(n, \Z_q^\ell, \Z_q^\ell, D_{\Z, \sigma})$ and $U(\Z_q^{\ell \times n} \times \Z_q^n)$. Then, by Theorem~\ref{thm:uniform-to-binary-secrets}, there is no distinguisher between
$\LWE(m, \Z_q^n, \{+1, -1\}^n, D_{\Z, \sigma_1})$ and $U(\Z_q^{n \times m} \times \Z_q^{m})$ with advantage $\epsilon$, where $\sigma_1 = 2 \sigma \sqrt{n + 1}$, and all other sufficient conditions are met by the hypotheses of the corollary. (From here on out, we omit additive $\negl(\secp)$ terms in the distinguishing probability for simplicity.)

Now, since the secrets all have fixed norm $\sqrt{n}$, we can apply Theorem~\ref{thm:fixed-norm-lwe-clwe}. Then, for parameters
\[ \gamma = \sqrt{n} \cdot \sqrt{\ln(m) + \ln(n) + \omega(\log \secp)} = O \left(\sqrt{n} \cdot \sqrt{\ln(m) + \omega(\log \secp)} \right),\]
and
\[\beta = O \left( \frac{\sigma_1}{q} \right) = O \left( \frac{\sigma \sqrt{n} }{q} \right),  \]
there is no distinguisher between $\CLWE(m, D_1^n, \frac{1}{\sqrt{n}} \{+1, -1\}^n, \gamma, \beta)$ and $D_1^{n \times m} \times U(\T^{m})$, as long as
\[ \sigma_1 = 2\sigma \sqrt{n+1} \geq 3 \sqrt{n} \sqrt{\ln(m) + \ln(n) + \omega(\log \secp)}, \]
which indeed holds as long as $\sigma \geq C \cdot \sqrt{\ln(m) + \omega(\log \secp)}$ for some universal constant $C$.

Lastly, we make the secret direction for the $\CLWE$ distribution a completely random unit vector in $\R^n$ via Lemma~\ref{lemma:wc-to-ac-clwe}. This has no effect on any of the parameters, so this means there is no distinguisher between $\CLWE(m, D_1^n, \gamma, \beta)$ and $D_1^{n \times m} \times U(\T^{m})$, as desired.
\end{proof}

\subsection{Hardness of Sparse CLWE}

In this subsection, we take advantage of our reduction from $\LWE$ to $k$-sparse $\LWE$ to reduce $\LWE$ to a $k$-sparse version of $\CLWE$ with a very similar proof to that of Corollary~\ref{cor:lwe-to-clwe}. Later on, the main benefit of this reduction is that in the resulting $\CLWE$ distribution, $\gamma$ will be small, which will result in a family of GMM instances, each with a small number of Gaussians.

\begin{corollary}[Reduction from LWE to $k$-sparse CLWE]
\label{cor:LWE-to-sparse-CLWE} \sloppy Suppose $\log(q)/2^\ell = \negl(\secp), \sigma \geq 2 \cdot \sqrt{\ln n + \ln m + \omega(\log \secp) }$, and $k \log_2(n/k) \geq (\ell + 1) \log_2(q) + \omega(\log \secp)$. Then, for parameters
\[ \gamma = \sqrt{k} \cdot \sqrt{\ln(m) + \ln(n) + \omega(\log \secp)} \]
and
\[ \beta =  O \left( \frac{\sigma \sqrt{k}}{q} \right) \]
for some universal constant $C$, if $\LWE(n, \Z_q^\ell, \Z_q^\ell, D_{\Z, \sigma})$ and $U(\Z_q^{\ell \times n} \times \Z_q^{n})$ have no $T + \poly(n, m, \log(q), \log(\sigma), \log(\secp))$ time distinguisher with advantage $\epsilon$, then $\CLWE(m, D_1^n, \frac{1}{\sqrt{k}} \S_{n,k}, \gamma, \beta)$ and $D_1^{n \times m} \times U(\T^m)$ have no $T$-time distinguisher with advantage $2\epsilon m + \negl(\secp)$.
\end{corollary}

\begin{proof}
By Corollary~\ref{cor:useful-lwe-to-k-sparse-lwe}, we know there is no distinguisher with advantage $2\epsilon m + \negl(\secp)$ between $\LWE(m, \Z_q^n, \S_{n,k}, D_{\Z, \sigma'})$ and $U(\Z_q^{n \times m} \times \Z_q^m)$, where $\sigma' = 2 \sigma \sqrt{k+1}$. Now, applying Theorem~\ref{thm:fixed-norm-lwe-clwe}, since all secret vectors have norm $\sqrt{k}$, for parameters
\[ \gamma = \sqrt{k} \cdot \sqrt{\ln(m) + \ln(n) + \omega(\log \secp)} \]
and
\[ \beta = O \left( \frac{\sigma'}{q} \right) = O \left( \frac{\sigma \sqrt{k}}{q} \right)  ,\]
there is no $T$-time distinguisher with advantage $2 \epsilon m + \negl(\secp)$ between $\CLWE(m, D_1^n, \frac{1}{\sqrt{k}} \S_{n,k}, \gamma, \beta)$ and $D_1^{n \times m} \times U(\T^m)$, as long as
\[ \sigma' = 2 \sigma \sqrt{k+1} \geq 3\sqrt{k} \sqrt{\ln(m) + \ln(n) + \omega(\log \secp)},\]
which indeed holds by our assumption on $\sigma$.
\end{proof}

\subsection{Classical Hardness of CLWE}

With our reduction from fixed-norm LWE to CLWE, we can now show that worst-case lattice problems reduce \emph{classically} to CLWE, whereas Corollary~3.2 of \cite{CLWE} gives a \emph{quantum} reduction from worst-case lattice problems to CLWE. This now essentially follows from the following theorem due to \cite{brakerski2013classical}:

\begin{theorem}[Theorem~1.1 of \cite{brakerski2013classical}, informal]\label{thm:clasical-hardness-lwe}
There is an efficient classical reduction from (worst-case) $\sqrt{n}$-dimensional $\mathsf{gapSVP}$ to decisional $\LWE$ in dimension $n$ with modulus $q = \poly(n)$.
\end{theorem}
Given Theorem~\ref{thm:clasical-hardness-lwe}, we can now prove Corollary~\ref{cor:classical-hardness-clwe}.

\begin{proof}[Proof Sketch of Corollary~\ref{cor:classical-hardness-clwe}]
One way to approach this (with slightly worse parameters) is to directly combine Theorem~\ref{thm:clasical-hardness-lwe} and Corollary~\ref{cor:lwe-to-clwe}. However, to be less wasteful, we briefly describe below how to optimize the reduction by bypassing LWE with $U(\Z_q^n)$ secrets and working instead with just binary secrets. In fact, Theorem~\ref{thm:clasical-hardness-lwe} uses a definition of LWE with continuous noise, so one has to be a bit careful regardless.

At a very high level, we combine Theorem~\ref{thm:clasical-hardness-lwe} and Theorem~\ref{thm:fixed-norm-lwe-clwe}, but modified (in a small way) so that the LWE distribution resulting from Theorem~\ref{thm:clasical-hardness-lwe} has fixed norm. We modify their reduction as follows:
\begin{itemize}
    \item We observe that their modulus switching reduction, Corollary~3.2, preserves the secret distribution $U(\{0, 1\}^n)$. The last step of their reduction, just after applying Corollary~3.2, reduces this secret distribution, $U(\{0, 1\}^n)$, to $U(\Z_q^n)$ by a standard random self-reduction. This has the the effect of going back from binary LWE to standard LWE to finish the reduction. In our case, we remove this final reduction and keep the secret distribution binary.
    \item Furthermore, throughout the reduction, we substitute $U(\{0,1\}^n)$ secrets with $U(\{+1, -1\}^n)$ secrets. To do this, we modify Theorem~4.1 in \cite{brakerski2013classical} to handle $U(\{+1, -1\}^n)$ secrets. Their proof of Theorem~4.1 is general in that it only requires the secret distribution to be efficiently samplable, have enough high min-entropy as needed to apply the leftover hash lemma, have norm at most $\sqrt{n}$, and have small ``quality'' (see Definition~4.5 of \cite{brakerski2013classical}). Since the quality of $U(\{+1, -1\}^n)$ can be bounded above by $2$, it is easy to see that $U(\{+1, -1\}^n)$ satisfies all of these conditions. (We note there are other ways to make this change; Theorem~\ref{thm:uniform-to-binary-secrets}, due to \cite{Mic18}, shows a reduction from $U(\Z_q^n)$ with $U(\{+1, -1\}^n)$ with very similar parameters. In fact, if $q$ is odd, $\{+1, -1\}^n$ secrets and $\{0, 1\}^n$ secrets have straightforward reductions to each other, as shown in Lemma~2.12 and Lemma~2.13 of \cite{Mic18}.) The only reason we make this change is that the secrets will now have fixed norm $\sqrt{n}$ (instead of norm at most $\sqrt{n}$), which will allow us to use our fixed-norm LWE to CLWE reduction. Lastly, Corollary~3.2 of \cite{brakerski2013classical} simply requires an upper bound on the norm of the secret distribution, so the same result holds for $U(\{+1, -1\}^n)$ secrets.
\end{itemize}

Therefore, we have a (classical) reduction from worst-case lattice problems in dimension $\sqrt{n}$ to (decisional) LWE in dimension $n$ with $q = \poly(n)$, with secret distribution $\S = U(\{+1, -1\}^n)$ and continuous Gaussian errors. Thus, we can just use Lemma~\ref{lemma:discrete-to-continuous-samples} and Lemma~\ref{lemma:uniform-to-gaussian-samples} to reduce to CLWE with $r = \sqrt{n}$. If desired, one can use Lemma~\ref{lemma:wc-to-ac-clwe} to make the secret distribution $U(S^{n-1})$ instead of $U(\frac{1}{\sqrt{n}} \{+1, -1\}^n)$. (The exact parameter dependencies come from combining Theorem~2.16 of \cite{brakerski2013classical}, Theorem~2.17 of \cite{brakerski2013classical}, Theorem~4.1 of \cite{brakerski2013classical}, Corollary~3.2 of \cite{brakerski2013classical}, Lemma~\ref{lemma:discrete-to-continuous-samples}, Lemma~\ref{lemma:uniform-to-gaussian-samples}, and optionally Lemma~\ref{lemma:wc-to-ac-clwe}.)
\end{proof}

%% file: gmm.tex
\section{Hardness of Density Estimation for Mixtures of Gaussians}\label{sec:gmm-hardness}

Now, using tools from the previous sections, we reduce LWE to density estimation for mixtures of Gaussians, using similar ideas as \cite{CLWE}. Our machinery from the previous sections now allows us to give a fine-grained version of hardness of learning mixtures of Gaussians.

\begin{lemma}[Reducing LWE to GMM via $k$-sparse CLWE]\label{lemma:lwe-to-gmm-sparse}
Suppose $\log(q)/2^\ell = o(1), \sigma \geq 10 \cdot \sqrt{\ln n + \ln m}$, $k \log_2(n/k) \geq (\ell + 1) \log_2(q) + \omega(1)$, $q = \omega(\sigma \sqrt{k})$ and $q \le m^2$. Then, for
\[g =  O \left( \sqrt{k \ln(m)} \cdot \sqrt{\ln(m) + \ln(n)} \right), \]
if $\LWE(n, \Z_q^\ell, \Z_q^\ell, D_{\Z, \sigma})$ and $U(\Z_q^{\ell \times n} \times \Z_q^{n})$ have no $T + \poly(n, m, q)$ time distinguisher with advantage $\Omega(1/m^3)$, then density estimation for GMM in dimension $n$ with $g$ Gaussian components and $m$ samples has no $T$-time solver.
\end{lemma}

\begin{proof}
In short, this follows by composing the reductions from LWE to $k$-sparse CLWE (Corollary~\ref{cor:LWE-to-sparse-CLWE}),  from CLWE to hCLWE (Lemma~\ref{lemma:clwe-to-hclwe}), and from hCLWE to density estimation for mixtures of Gaussians (Theorem~\ref{thm:hclwe-to-gmm}).

As used in Corollary~\ref{cor:LWE-to-sparse-CLWE}, let $\beta = \Theta(\sigma \sqrt{k}/q) = o(1)$, and let $m'$ denote the number of CLWE samples. In anticipation of applying Lemma~\ref{lemma:clwe-to-hclwe} in reducing CLWE to hCLWE with $\delta = \beta$, we set
\begin{align*}
    m' = \Theta\left(\frac{m}{\beta}\right) = \Theta\left(\frac{mq}{\sigma \sqrt{k}}\right) < m^3,
\end{align*}
where the final inequality holds (for, say, sufficiently large values of $m$) since $q \le m^2$, $k \geq 1$, and $\sigma \geq 10 \sqrt{\ln n + \ln m} = \omega(1)$. Since $m' < m^3$, it follows that $\ln(m') < \ln(m^3) = 3 \ln(m)$. 

To reduce LWE to $k$-sparse CLWE, we apply Corollary~\ref{cor:LWE-to-sparse-CLWE} with $\epsilon = 1/(6m')$. Since we have the conditions $\log(q)/2^\ell = o(1)$, $k \log_2(n/k) \geq (\ell + 1) \log_2(q) + \omega(1)$, and
\begin{align*}
    \sigma \ge 10 \sqrt{\ln n + \ln m} > 3 \sqrt{\ln n + \ln m'} > 2\sqrt{\ln n + \ln m' + \omega(1)},
\end{align*}
one can choose sufficiently small $\secp = \omega(1)$ to satisfy the conditions of Corollary~\ref{cor:LWE-to-sparse-CLWE} such that the $\negl(\secp)$ additive term in the advantage loss is at most $1/100$ and such that
\[
\gamma = \sqrt{k} \cdot \sqrt{\ln(m') + \ln(n) + \omega(\log \secp )} = O \left( \sqrt{k} \cdot \sqrt{\ln(m) + \ln(n)} \right).
\]
Corollary~\ref{cor:LWE-to-sparse-CLWE} then implies that there is no $T$-time distinguisher with advantage
\[ 2 \epsilon m' + \frac{1}{100} < \frac{2}{5}\]
between $\CLWE(m', D_1^n, \frac{1}{\sqrt{k}} \S_{n,k}, \gamma, \beta)$ and $D_1^{n \times m'} \times U \left(\T^{m'} \right)$. By Lemma~\ref{lemma:clwe-to-hclwe}, we reduce $m'$ samples of $\CLWE$ to $m$ samples of $\hCLWE$ with parameter $\delta = \beta$, so that $\beta' = \sqrt{2} \beta$ and $\gamma' = \gamma$, at the cost of $\poly(n, m, 1/\beta) = \poly(n, m, q)$ time and $1/1000$ additional failure probability. Then, by Theorem~\ref{thm:hclwe-to-gmm}, there is no GMM learner for
\[ g = 4 \gamma \sqrt{\ln(m)/\pi} + 1 = O \left( \sqrt{k \ln(m)} \cdot \sqrt{\ln(m) + \ln(n)} \right) \]
Gaussians, as long as $\beta' < 1/32$, which holds in our case as $\beta' = \sqrt{2} \beta = o(1)$.
\end{proof}

Now, we set parameters and invoke Lemma~\ref{lemma:lwe-to-gmm-sparse}.

\begin{corollary}\label{cor:lwe-to-gmm-parameterized-general}
Suppose $10 \sqrt{\ln(m) + \ln(n)} \leq \sigma$, $\omega(\sigma \sqrt{k}) \leq q \leq \poly(\ell)$, $k \log_2(n/k) = (1 + \Theta(1)) \ell \log_2(q)$, $q \le m^2$, and $m \leq \poly(n)$, and suppose that $\LWE(n, \Z_q^\ell, \Z_q^\ell, D_{\Z, \sigma})$ and $U(\Z_q^{\ell \times n} \times \Z_q^n)$ have no $T(\ell) + \poly(n)$ time distinguisher with advantage at least $\Omega(1/m^3)$. Then, there is no algorithm solving density estimation in dimension $n$ with $m$ samples for $g$ Gaussians, where
\[g = O\left( \sqrt{k \cdot \log(m) \cdot \log(n)} \right). \]
\end{corollary}

\begin{proof}
First, since $q \le \poly(\ell)$, we have $\log(q) / 2^\ell \le O(\log(\ell)/2^\ell) = o(1)$. Thus, we can invoke Lemma~\ref{lemma:lwe-to-gmm-sparse}. This gives
\[ g = O\left( \sqrt{k \ln (m)} \cdot \sqrt{\ln(m) + \ln(n)} \right) = O\left( \sqrt{k \cdot \log(m) \cdot \log(n)}\right), \]
as $\ln(m) = O(\log n)$ by our assumption that $m \le \poly(n)$.
\end{proof}

\begin{corollary}\label{cor:main-result}
Let $\epsilon, \delta \in (0,1)$ be arbitrary constants with $\delta < \epsilon$ and let $n = 2^{\ell^{\delta}}$. Assuming
\[\LWE \left( 2^{\ell^\delta}, \Z_q^{\ell}, \Z_q^{\ell}, D_{\Z, \sigma} \right)\]
has no $2^{O(\ell^{\epsilon})}$ time distinguisher from $U\left( \Z_q^{\ell \times 2^{\ell^\delta}} \times \Z_q^{2^{\ell^\delta}} \right)$ with advantage at least $\Omega \left( m \left( 2 ^ {\ell^{\delta}} \right)^{-3} \right)$, where $\sigma = \ell^{1/2}$ and $q = \ell^2$, then there is no algorithm solving density estimation for $g$ Gaussians in $\R^n$ with $m = m(n)$ samples in time $2^{\log_2(n)^{\epsilon/\delta}}$, where $\ell \le m(n) \le \poly(n)$ and $g = O\left( (\log n)^{1/(2 \delta)} \cdot \sqrt{\log (m(n))} \cdot \sqrt{\log \log n} \right)$.

In particular, for the number of GMM samples $m(n)$ satisfying $\ell \le m(n) \le \poly(\log(n)) = \poly(\ell)$, we have $g = O((\log n)^{1/(2 \delta)} \cdot \log \log n)$ assuming there is no $1/\poly(\ell)$-advantage distinguisher for LWE, and for $m(n)$ satisfying $\ell \le m(n) = \poly(n) = 2^{O(\ell^{\delta})}$, we have $g = O((\log n)^{1/2 + 1/(2 \delta)} \cdot \sqrt{\log \log n})$ assuming there is no $1/2^{O(\ell^{\delta})}$-advantage distinguisher for LWE.
\end{corollary}

\begin{proof}
We set $n = 2^{\ell^\delta}$ and $k = 4 \ell^{1 - \delta} \log_2(\ell)$ in Corollary~\ref{cor:lwe-to-gmm-parameterized-general}. Let us first confirm that all the hypotheses of Corollary~\ref{cor:lwe-to-gmm-parameterized-general} hold. First, observe that 
\[ 10 \sqrt{\ln n + \ln m} = O(\ell^{\delta/2}) = o(\ell^{1/2}) \leq \sigma.\]
We also have $q = \ell^2 = \omega(\ell^{1/2} \cdot \ell) \geq \omega(\sigma \cdot \sqrt{k})$, as needed. Further, we have
\[ k \log_2(n/k) = (4 - o(1)) \ell^{1 - \delta} \log_2(\ell) \ell^{\delta} = (4 - o(1)) \ell \log_2(\ell) = (2 - o(1)) \ell \log_2(q),\]
and lastly, $q = \ell^2 \le m^2$, as needed. If we have a 
\[ 2^{\ell^\epsilon} = 2^{\log_2(n)^{\epsilon/\delta}} \]
time distinguisher for the mixture of Gaussians, we get a $2^{\ell^\epsilon} + \poly(n) = 2^{O(\ell^{\epsilon})}$ time algorithm for $\LWE$. The number of Gaussians becomes
\begin{align*} g = O \left(\sqrt{k \cdot \log m \cdot \log n} \right) &= O\left( \sqrt{ \ell^{1 - \delta} \cdot \log(\ell) \cdot \log(n) \cdot \log(m)} \right)
\\&= O\left( (\log_2 n)^{\frac{1 - \delta}{2 \delta}} \cdot \sqrt{\log \log n} \cdot \sqrt{\log n} \cdot \sqrt{\log m} \right)
\\&= O\left( (\log n)^{\frac{1}{2 \delta}} \cdot\sqrt{\log m} \cdot \sqrt{\log \log n} \right),
\end{align*}
as desired.
\end{proof}

We give another setting of parameters where the number of Gaussian components in the mixture is larger, but assumption on LWE is weaker.

\begin{corollary}\label{cor:main-result-poly-lwe}
Let $\alpha > 1$ be an arbitrary constant. Assuming $\LWE(n, \Z_q^\ell, \Z_q^\ell, D_{\Z, \sigma})$ and $U(\Z_q^{\ell \times n} \times \Z_q^{n})$ have no $T(\ell) + \poly(n)$ time distinguisher with advantage $\Omega(1/m^3)$ where $n = \ell^{\alpha}$, $\sigma = \ell^{1/2}$ and $q = \ell^2$, then there is no algorithm solving density estimation for mixtures of $g$ Gaussians with $m$ samples in time $T(\ell) = T(n^{1/\alpha})$, where $g = O\left( n^{1/(2\alpha)} \cdot \log n \right)$ and $\ell \le m \le \poly(n) = \poly(\ell)$.

In particular, if $T(\ell) = \poly(\ell)$, then assuming the LWE problem is hard to distinguish for $\poly(\ell)$-time algorithms with advantage $1/\poly(\ell)$, then density estimation cannot be solved in $\poly(n)$ time with $\poly(n) \ge \ell$ samples for $g = n^{\Omega(1)}$ Gaussians.
\end{corollary}

\begin{proof}
We set $k= 4 \ell / (\alpha-1) = 4 n^{1/\alpha}/(\alpha-1)$ and apply Corollary~\ref{cor:lwe-to-gmm-parameterized-general}. Observe that
\begin{align*} 
k \log_2(n/k) = \frac{4 \ell}{\alpha-1} \cdot \log_2 \left( \frac{\ell^{\alpha}}{4 \ell /\alpha} \right) &= \frac{4 \ell}{\alpha-1} \cdot ( (\alpha - 1) \log_2(\ell) - O(1)) 
\\&= 4 \ell \log_2(\ell) - O(\ell)
\\&= 2 \ell \log_2(q) - O(\ell)
\\&= (1 + \Theta(1)) \ell \log_2(q),
\end{align*}
as necessary. Let us see that the other hypotheses of Corollary~\ref{cor:lwe-to-gmm-parameterized-general} hold. We have
\[ 10 \sqrt{\ln n + \ln m} = O \left( \sqrt{\log \ell} \right) = o(\sigma).\]
Also observe that $q = \ell^2 \geq \omega(\ell^{1/2} \cdot \ell) \geq \omega (\sigma \cdot \sqrt{k})$ and $q = \ell^2 \le m^2$.

If we have a time $T(n^{1/\alpha}) = T(\ell)$ distinguisher for hCLWE, we get a time $T(\ell) + \poly(n)$ time distinguisher for LWE. The number of Gaussian components becomes
\[ g = O\left( \sqrt{k \cdot \log(m) \cdot \log(n)} \right) = O \left( n^{1/(2\alpha)} \cdot \log(n) \right). \]
\end{proof}

%% file: alternate.tex
\section{Alternate Reduction from LWE to CLWE}\label{appendix:alternate-lwe-to-clwe}

In this section, we propose an alternate reduction from LWE to CLWE than that of Corollary~\ref{cor:lwe-to-clwe}. We note that we reduce to \emph{search} CLWE, and not decisional CLWE. Here is a brief outline to the steps of this reduction:

\begin{enumerate}
    \item First, we start with the standard search version of LWE (dimension $n$, mod $q$, and noise $D_{\Z, \sigma}$). 
    \item Then, we reduce to the (search) ``Hermite normal form'' of LWE, where the secret is drawn from the \emph{error distribution} instead of uniform over $\Z_q^n$ (with a small additive blowup in the number of samples), following \cite{applebaum2009fast, micciancio2009lattice} (and the more refined analysis by \cite{brakerski2013classical}). 
    \item Since the secrets are now short, we know there is some (small) $r \approx \sigma \sqrt{n}$ for which non-negligibly often, secrets will have $\ell_2$ norm \emph{exactly} $r$. The reduction in this step is the trivial reduction, but crucially uses the fact that this is a \emph{search} reduction.
    \item Since the secrets now have fixed (and small) norm, we use Theorem~\ref{thm:fixed-norm-lwe-clwe} to reduce to CLWE, slightly modified to be a search reduction (as opposed to a decision reduction).
\end{enumerate}

Explicitly, we have the following theorem.

\begin{theorem}[Alternate Reduction from LWE to CLWE]
Suppose there exists no algorithm running in time $T + \poly(n,m, \log(\secp), \log(q))$ that outputs $\vecs$ with probability $\epsilon$ when given $(A, \vecs^\top A + \vece \pmod{q})$, where $A \sim U(\Z_q^{n \times m})$, $\vecs \sim U(\Z_q^n)$, and $\vece \sim D_{\Z^m, \sigma}$. Suppose $q \leq 2^{2^{O(n)}}$ and $\sigma \geq 2 \sqrt{\ln(n)}$. Then, there is no $T$-time algorithm outputting $\vecs'$ with probability at least $(\epsilon + 2^{-n})\cdot 2 \sigma^2  n + \negl(\secp)$ when given $(A', \gamma \cdot (\vecs')^{\top} A' + \vece \pmod{1})$, where $A' \sim (D_1)^{n \times m'}$,  $\vecs' \sim \frac{1}{r} \S$, $\vece \sim D_{\beta}^{m'}$, where $\S$ is the set of all vectors in $\Z^n$ with norm exactly $r$, for some $r = O(\sigma \sqrt{n})$, and where
\begin{align*}
    m' &= m + O(n),
    \\\gamma &= O\left( \sigma \sqrt{n} \cdot \sqrt{\ln(m) + \ln(n) + \omega(\log \secp)} \right),
    \\\beta &= O\left( \frac{\sigma \sqrt{n} \cdot \sqrt{\ln(m) + \ln(n) + \omega(\log \secp)}}{q} \right).
\end{align*}
\end{theorem}

Note that one can reduce to secret distribution $\vecs' \sim U(S^{n-1})$ if desired, by using (a search version of) Lemma~\ref{lemma:wc-to-ac-clwe}.

\begin{proof}
First, we invoke Lemma~2 of \cite{applebaum2009fast} to turn the secret distribution from $\Z_q^n$ into $D_{\Z^n, \sigma}$. The only loss in parameters we get is the number of samples, which becomes $m' = m + O(n) + O(\log \log q)$, as $O(n) + O(\log \log q)$ samples suffice to efficiently find $n$ linearly independent vectors over $\Z_q$, with probability of failure at most $2^{-n}$. (See Claim 2.13 of the full version of \cite{brakerski2013classical}.) Since $q \leq 2^{2^{O(n)}}$, $O(\log \log q)$ can be absorbed into $O(n)$, making $m' = m + O(n)$. This means there is no solver for $\vecs$ with probability at least $\epsilon + 2^{-n}$ for this ``normal form'' of LWE (i.e. with $\vecs \sim D_{\Z^n, \sigma}$).
 
Now, we observe that there exists some $r \leq \sigma \sqrt{n}$ such that non-negligibly often, secrets $\vecs \sim D_{\Z^n, \sigma}$ will have $\ell_2$ norm exactly $r$. To see this, we use the proof of Lemma~4.4 in \cite{micciancio2007worst} (ultimately based on Lemma~1.5 of \cite{banaszczyk1993new}) to see that the probability that $\norm{\vecs} \geq \sigma \sqrt{n}$ is at most $2^{-n}$ for $\vecs \sim D_{\Z^n, \sigma}$. Conditioned on $\norm{\vecs} \leq \sigma \sqrt{n}$, since $\norm{\vecs}^2$ is a non-negative integer, we know it must take on at most $\sigma^2 \cdot n+1$ different values. Let $\S_r$ be the set of all vectors in $\Z^n$ with $\ell_2$ norm exactly $r \in \R$. What we have just shown is that there exists some $r \leq \sigma \sqrt{n}$ such that $\Pr_{\vecs \sim D_{\Z^n, \sigma}}[\vecs \in \S_r] \geq (1 - 2^{-n})/(\sigma^2 \cdot n + 1) \geq 1/(2 \sigma^2 n)$. From here on out, we now fix $r$ to be such an $r$. Moreover, it is easy to see that $r \geq \sigma$, as the only way for the norm to be below $\sigma$ is if all coordinates of the discrete Gaussian have magnitude at most $\sigma$, which happens with exponentially small probability in $n$.

Therefore, if we have some solver with success probability $(\epsilon + 2^{-n}) \cdot 2 \sigma^2 n$ for LWE with $\vecs \sim \S$, then that same solver has success probability at least $\epsilon + 2^{-n}$ for LWE with $\vecs \sim D_{\Z^n, \sigma}$. Therefore, we now know there is no solver for $\LWE$ with secret distribution $\S$ with success probability $(\epsilon + 2^{-n}) \cdot 2 \sigma^2 n$.

Before invoking Theorem~\ref{thm:fixed-norm-lwe-clwe}, we simply increase the width of the noise, as the requirement on the width of the noise is large for the reduction to go through. Specifically, we set $\sigma' = 3 r \sqrt{\ln(m') + \ln(n) + \omega(\log \secp)} \geq \sigma$, where the inequality comes from the fact that $r \geq \sigma$. We can achieve this reduction by simply adding noise.

Now, we directly invoke Theorem~\ref{thm:fixed-norm-lwe-clwe}, as our requirement on $\sigma'$ is now satisfied. While the reduction is formally a decisional reduction, the proof also works in the search setting. In fact, throughout the reduction, the secret remains the same, up to scaling by $r$. This implies there is no solver with success probability at least $(\epsilon + 2^{-n}) \cdot 2 \sigma^2 n + \negl(\secp)$, for parameters
\begin{align*}
    \gamma &= r \cdot \sqrt{\ln(m') + \ln(n) + \omega(\log \secp)} \leq O \left( \sigma \sqrt{n} \cdot \sqrt{\ln(m) + \ln(n) + \omega(\log \secp)} \right),
    \\\beta &= O \left( \frac{\sqrt{(\sigma')^2 + r^2(\ln(m') + \ln(n) + \omega(\log \secp)}}{q} \right) = O \left( \frac{\sigma \sqrt{n} \cdot \sqrt{\ln(m) + \ln(n) + \omega(\log \secp)}}{q} \right).
\end{align*}
\end{proof}

%% file: lowsample-alg.tex
\section{Low-Sample Algorithm for Sparse \texorpdfstring{$\hCLWE$}{hCLWE}}
\newcommand{\A}{\hCLWE^{(g)}(m, D_1^n, \frac{1}{\sqrt{k}} \S_{n,k}, \gamma, \beta)}

\begin{theorem}\label{thm:algorithm-brute-force-hclwe}
\sloppy Let $m = 5 k \log_2(n) / \log_2(1 / (\beta \sqrt{k}))$. Suppose $\gamma \geq 2\sqrt{k(\ln n + \ln m)}$ and $\log_2(1/(\beta \sqrt{k})) = \omega(\log \log m)$. Then, there is a $O\left(m \cdot \poly(n) \cdot 2^k \binom{n}{k} \right)$-time algorithm using $m$ samples that learns the parameters for GMM when restricted to ($m$ sample) mixtures $D_1^{n \times m}$ and $\hCLWE^{(g)}(m, D_1^n, \frac{1}{\sqrt{k}} \S_{n,k}, \gamma, \beta)$ for any fixed $g \geq C \cdot \gamma \cdot \sqrt{\log m}$ for some universal constant $C$. That is, we view all the parameters as fixed, and the algorithm either learns the correct secret $\vecs \sim \frac{1}{\sqrt{k}} \S_{n,k}$ (and thus the corresponding $\hCLWE^{(g)}$ distribution), or knows that the distribution is $D_1^n$, with success probability at least $9/10$ in both cases.
\end{theorem}

\begin{remark}
This theorem can be generalized for other settings of $\beta, \gamma$, but we state it this way because it suffices for our purposes. It also works for the setting of non-truncated hCLWE.
\end{remark}

\begin{remark}
While the runtime of this algorithm is similar to the algorithm solving hCLWE given in Theorem~7.5 of \cite{CLWE} as applied in a black-box way, the sample complexity needed here is $\ll k \log_2(n)$, as opposed to roughly $2^{O(\gamma^2)} = n^{\Omega(k)}$. 
\end{remark}

\begin{algorithm}
\SetKwInOut{Input}{Input}
\SetKwInOut{Output}{Output}
\BlankLine
\textbf{Input:} Sampling oracle to distribution $\mathcal{D}$.

\textbf{Output:} $\vecs$ to indicate $\mathcal{D} = \hCLWE^{(g)}$ with secret $\vecs$, and $0$ for $\mathcal{D} = D_1^n$.
\BlankLine

Draw $m$ samples $\veca_1, \ldots, \veca_m \sim \mathcal{D}$.

\For{$\vecs \in \frac{1}{\sqrt{k}} \S_{n,k}$}{
Compute $f_\vecs(\veca_i) = \inner{\veca_i, \vecs} \mod \gamma/(\sqrt{k} \cdot \gamma'^2)$ for all $i \in [m]$.

\If{$f_\vecs(\veca_i) \in [-a \beta/\gamma', a \beta/\gamma']$ for all $i \in [m]$}{
\KwRet{$\vecs$}.
}
}
\KwRet{$0$}.
\BlankLine

\caption{Low Sample algorithm for $\hCLWE^{(g)}$}\label{}
\end{algorithm}

\begin{proof}
Let $t := |\S_{n, k}| = \binom{n}{k} \cdot 2^k$ denote the number of $k$-sparse $\{-1, 0, +1\}$-secrets. For the sake of this proof, we take the representatives of $\T_q$ to be in the interval $[-q/2, q/2)$. Further, let $\gamma' = \sqrt{\gamma^2 + \beta^2}$ and $\veca \in \R^n$ and $\vecs \in \frac{1}{\sqrt{k}}\S_{n,k}$. We define $f_{\vecs} : \R^n \to \T_{\gamma / (\sqrt{k} \cdot \gamma'^2)}$ by
\begin{align*}
   f_\vecs(\veca) := \inner{\veca, \vecs} \mod \gamma/(\sqrt{k} \cdot \gamma'^2).
\end{align*}
We use the main idea in the proof of Claim 5.3 in \cite{CLWE} to give an algorithm that finds the correct secret $\vecs$, if it exists, or report that none exists if the distribution is $D_1^n$. Given $m$ samples $\veca_1, \ldots, \veca_m$ from an unknown distribution $\mathcal{D}$, we compute $f_{\vecs}(\veca_i)$ for all possible secret directions $\vecs \in \frac{1}{\sqrt{k}} \S_{n, k}$ and for all samples $i \in [m]$. This takes time $O(m \cdot t \cdot \poly(n))$, where we allow $\poly(n)$ time to take numbers mod $\gamma / (\sqrt{k} \gamma'^2)$. If there is some $\vecs$ such that $f_\vecs(\veca_i)$ is small for all samples $i \in [m]$, then we output $\vecs$, and otherwise we guess $\mathcal{D} = D^m_1$.

Now, suppose that the input distribution is $\mathcal{D} = \hCLWE^{(g)}(m, D_1^n, \frac{1}{\sqrt{k}} \S_{n,k}, \gamma, \beta)$. Let $\vecs^*$ be the randomly sampled but fixed secret direction. Then for all the $m$ samples $\veca_i$, we have that $\inner{\vecs^*, \veca_i} \pmod{\gamma / (\gamma'^2)}$ is distributed as $D_{\beta/\gamma'} \mod \gamma/\gamma'^2$. This can be seen from Equation~\ref{eqn:truncated-hclwe-pdf}. (As an aside, note that by Claim 5.3 of \cite{CLWE} this holds even when the input distribution is not truncated, that is, $\mathcal{D} = \hCLWE(m, D_1^n, \frac{1}{\sqrt{k}} \S_{n,k}, \gamma, \beta)$.) Now, supposing for simplicity that $k$ is a perfect square, we can take this mod $\gamma / (\sqrt{k} \cdot \gamma'^2)$ to get that $f_{\vecs^*}(\veca_i)$ is distributed as $D_{\beta/\gamma'} \mod \gamma/(\sqrt{k} \cdot \gamma'^2)$. (In case $k$ is not a perfect square, we can take $\gamma/\left( \ceil*{\sqrt{k}} \cdot \gamma'^2 \right)$ as the modulus instead.)

For a parameter $\delta > 0$ specified later, let $a = \sqrt{\ln(1/\delta)}$. By a standard Chernoff bound, the probability mass of $D_{\beta/\gamma'}$ that is outside the interval $[-a \beta/\gamma', a \beta/\gamma']$ is at most $\delta$. Taking a union bound over the $m$ samples $\veca_i$,
\begin{equation}\label{eq:low-sample-true-secret}
    \Pr \left[ \exists i \in [m] \text{ s.t. } f_{\vecs^*}(\veca_i) \notin [-a \beta/\gamma', a \beta/\gamma'] \right] \leq m \delta = \frac{1}{100},
\end{equation}
when setting $\delta = 1/(100 m)$.

We still have to argue that for this $\mathcal{D}$ that no other $\vecs \neq \vecs^*$ passes the test. To see this, fix some $\vecs \neq \vecs^*$. Let $\vecz = \vecs - \vecs^* \in \frac{1}{\sqrt{k}} \cdot \{-2,-1,0,1,2\}^n$. Since $\vecs \in \frac{1}{\sqrt{k}} \S_{n,k}$ and $\vecs \neq \vecs^*$, it follows that $\norm{\vecz} \geq \sqrt{2}/\sqrt{k}$. We have 
\begin{equation}\label{eq:low-sample-expand-secret}
\inner{\veca_i, \vecs} = \inner{\veca_i, \vecs^* + \vecz} = \inner{\veca_i, \vecs^*} + \inner{\veca_i, \vecz}.
\end{equation}
On its own, for fixed $\vecz$, $\inner{\veca_i, \vecz}$ is distributed according to $D_{\norm{\vecz}}$, which is from a discrete Gaussian wider than $D_{1/\sqrt{k}}$. By Lemma~\ref{lemma:gaussian-mod-lattice-looks-uniform} (and the union bound over $m$ samples), it follows that the distribution of $(\inner{\veca_i, \vecz} \pmod{\gamma / (\sqrt{k} \cdot \gamma'^2)})_{i \in [m]}$ is $m \cdot \exp(- \gamma'^4/\gamma^2)/2$-close to $U(\T_{\gamma/(\sqrt{k} \cdot \gamma'^2)})^m$. Therefore,
\begin{align*}
    \Pr_{\veca_i} \left[\left(f_\vecz(\veca_i)\right)_{i \in [m]} \in [-2a \beta/\gamma', 2a \beta/\gamma']^m \right] &\le \Delta \left(D_1^m \mod \gamma/(\sqrt{k} \gamma'^2), U\left(\T_{\gamma / (\sqrt{k} \gamma'^2)}\right)^m\right) + \left( 4 a \beta \sqrt{k} \cdot \frac{\gamma'}{\gamma} \right)^m\\
    &\le m \exp(-\gamma^2)/2 + \left( 4 a \beta \sqrt{k} \cdot \frac{\gamma'}{\gamma} \right)^m\\
    &\le m \exp(-\gamma^2)/2 + \left( 8 a \beta \sqrt{k} \right)^m.
\end{align*}
Since this was for a particular secret $\vecs \neq \vecs^*$, we can union bound over all $\vecs \neq \vecs^*$ to see that
\begin{equation}\label{eq:low-sample-wrong-secret}
    \Pr_{\veca_i} \left[\exists \vecs \neq \vecs^* \text{ s.t. } \left(f_\vecz(\veca_i)\right)_{i \in [m]} \in [-2a \beta/\gamma', 2a \beta/\gamma']^m \right] \le t \cdot m \exp(-\gamma^2)/2 + t \cdot \left( 8 \cdot a \beta \sqrt{k} \right)^m.
\end{equation}
Note that if $f_{\vecz}(\veca_i) \notin [-2a \beta/\gamma', 2a \beta/\gamma']$ and $f_{\vecs^*}(\veca_i) \in  [-a \beta/\gamma', a \beta/\gamma']$, then by equation~\eqref{eq:low-sample-expand-secret}, it follows that $f_{\vecs}(\veca_i) \notin  [-a \beta/\gamma', a \beta/\gamma']$. Thus, equations~\eqref{eq:low-sample-true-secret} and~\eqref{eq:low-sample-wrong-secret} fully characterize the ``bad'' events, as if both events do not happen, then $\vecs^*$ passes the test, and no other $\vecs \neq \vecs^*$ passes the test. Therefore, if samples are from the hCLWE distribution, then the probability of failure is at most
\[ \frac{1}{100} + t \cdot m \exp(-\gamma^2)/2 + t \cdot \left( 8 a \beta \sqrt{k} \right)^m .\]
We first analyze the middle term. Since $\gamma \ge 2 \sqrt{k (\ln n + \ln m)}$, we have
\[ t \cdot m \cdot \exp(-\gamma^2)/2 \leq \frac{2^k \cdot n^k \cdot m}{2 \cdot m^{4k} \cdot n^{4k}} < \frac{1}{100} \]
for large enough $m$ and $k$. For the last term, we have
\begin{align*}
t \cdot \left( 8 a \beta \sqrt{k} \right)^m = t \cdot \left( 8 \cdot \ln(100m) \cdot \beta \sqrt{k} \right)^m &= t \cdot \left( 2^{3 + \log_2(\ln(100m)) - \log_2(1/(\beta \sqrt{k}))} \right)^m
\\&\leq t \cdot 2^{ - (m/2) \cdot \log_2(1/(\beta \sqrt{k}))}
\\&\leq 2^{k + k \log_2(n) - (m/2) \cdot \log_2(1/(\beta \sqrt{k}))}
\\&\leq 2^{-k \log_2(n)}
\\&\leq \frac{1}{100}
\end{align*}
for sufficiently large $k, n$, where we have used our hypothesis on $\log_2(1/(\beta \sqrt{k}))$ and choice of $m$. Thus, for the hCLWE distribution, we output the correct secret $\vecs^*$ with probability at least $19/20$.

Now, suppose we are given samples from $D_1^n$. For any fixed $\vecs \in \frac{1}{\sqrt{k}}\S_{n,k}$, we have $\inner{\veca_i, \vecs} \sim D_1$, independently of $\vecs$. By Lemma~\ref{lemma:gaussian-mod-lattice-looks-uniform} and Lemma~\ref{lemma:smoothing-parameter-estimate},
\begin{align*}
    \Delta(D_1^m \mod \gamma/(\sqrt{k} \cdot \gamma'^2), \T_{\gamma / (\sqrt{k} \cdot \gamma'^2)}^m ) \le m \exp(-\gamma'^4 k/\gamma^2)/2 \le m \exp(-\gamma^2)/2.
\end{align*}
Therefore, by a simpler analysis than the one above, we have
\begin{equation}
    \Pr_{\veca_i} \left[\exists \vecs \in \S_{n,k} \text{ s.t. } \left(f_\vecs(\veca_i)\right)_{i \in [m]} \in [-a \beta/\gamma', a \beta/\gamma']^m \right] \le t \cdot m \exp(-\gamma^2)/2 + t \cdot \left( 8 \cdot a \beta \sqrt{k} \right)^m,
\end{equation}
which we have previously bounded above by $2/100$. This completes the proof, as we will output 0 (to indicate $\mathcal{D} = D_1^n$) with probability at least $19/20$ in this case.
\end{proof}

Now, we combine Theorem \ref{thm:algorithm-brute-force-hclwe} and Corollary \ref{cor:main-result} to get the following tightness for the mixtures of Gaussians we consider.

\begin{corollary}\label{cor:gmm-upper-bound}
Following the notation of Corollary~\ref{cor:main-result}, there is an algorithm solving for the parameters for GMM, when restricted to $D_1^n$ and hCLWE, using $m = O(\ell) = O\left( (\log n)^{1/\delta} \right)$ samples and time $2^{O \left((\log n)^{1/\delta} \log \log n \right) }$.
\end{corollary}

\begin{proof}
We apply Theorem~\ref{thm:algorithm-brute-force-hclwe}. If we trace $\beta$ in the proof of Corollary~\ref{cor:main-result}, we see that
\[ \beta \sqrt{k} = O \left( \frac{\sigma k}{q} \right)  = O \left( \frac{\sqrt{\ell} \cdot \ell }{\ell^2}\right) = O\left(\ell^{-1/2} \right). \]
Therefore, $\log(1/(\beta \sqrt{k})) = \Omega(\log \ell)$, which implies
\[m = \frac{5k \log_2(n)}{\log_2(1/\beta \sqrt{k})} = O \left(\frac{\ell^{1 - \delta} \cdot \log(\ell) \cdot \ell^{\delta}}{\log (\ell)} \right) = O(\ell) = O((\log n)^{1/\delta}), \]
and thus that $\log(1/(\beta \sqrt{k})) = \omega(\log \log m)$. For the runtime, observe that
\begin{align*}
m \cdot \poly(n) \cdot 2^k \binom{n}{k} \leq m \cdot n^{O(k)} \leq m \cdot 2^{O(\log(n) \ell^{1 - \delta} \log(\ell))} &\leq \poly(\log n) \cdot 2^{O(\log(n)^{1/\delta} \log \log n)}
\\&=2^{O(\log(n)^{1/\delta} \log\log n)},
\end{align*}
as desired.
\end{proof}

%% file: CLWEtoLWE.tex
\section{Reduction from CLWE to LWE}\label{appendix:clwe-to-lwe}

Here, we show a reversed version of Theorem~\ref{thm:fixed-norm-lwe-clwe}, i.e. a reduction from discrete-secret CLWE to fixed-norm LWE. Note that this gives a reduction only from discrete-secret CLWE to LWE, and not CLWE with secrets $\vecs \sim U(S^{n-1})$ to LWE.

\begin{theorem}[CLWE to LWE]\label{thm:clwe-to-lwe}
Let $r \in \R_{\geq 1}$, and let $\S$ be an arbitrary distribution over $\Z^n$ where all elements in the support of $\S$ have $\ell_2$ norm $r$. Then, for 
\begin{align*}
\gamma &= r \cdot \sqrt{\ln(m) + \ln(n) + \omega(\log \secp)}, \text{  and}
\\\sigma &= O (\beta \cdot q),
\end{align*}
if there is no $T + \poly(n, m, \log(q), \log(\secp))$ time distinguisher between $\CLWE(m, D_1^n, \frac{1}{r} \cdot \S, \gamma, \beta)$ and $D_1^{n \times m} \times U(\T^m)$ with advantage at least $\epsilon - \negl(\secp)$, then there is no $T$-time distinguisher between $\LWE(m, \Z_q^n, \S, D_{\sigma})$ and $U(\Z_q^{n \times m} \times \T_q^m)$ with advantage $\epsilon$, as long as $\beta \cdot q \geq 3r \sqrt{\ln(m) + \ln(n) + \omega(\log \secp)}$.
\end{theorem}

Note that we reduce to a continuous-error version of LWE. Using standard techniques (see Theorem~3.1 of \cite{peikert2010efficient}), this can be reduced to discrete Gaussian errors. Similarly, the final LWE distribution can be made to have secrets $U(\Z_q^n)$ by a standard random self-reduction. Lastly, the proof of Theorem~\ref{thm:clwe-to-lwe} preserves the secret vector up to scaling, so it also is a reduction between the search versions of the problems.

With this reduction from discrete-secret CLWE to LWE, we get a search-to-decision reduction for discrete-secret CLWE. This can be obtained immediately by combining (the search version of) Theorem~\ref{thm:clwe-to-lwe}, standard search-to-decision reductions for $\LWE$ (see \cite{Peikert09, applebaum2009fast,MM11,micciancio2012trapdoors,brakerski2013classical}), and Theorem~\ref{cor:lwe-to-clwe} (or Theorem~\ref{thm:fixed-norm-lwe-clwe} if the $\LWE$ search-to-decision reduction preserves the norm of the secret). We leave open the question of whether there is a more direct search-to-decision reduction for $\CLWE$.

The steps of this proof are essentially just versions of Lemma~\ref{lemma:uniform-to-gaussian-samples} and Lemma~\ref{lemma:discrete-to-continuous-samples} but in the reverse directions. We give these ``reversed'' lemmas below.

\begin{lemma}[Reverse of Lemma~\ref{lemma:uniform-to-gaussian-samples}]\label{lemma:gaussian-to-uniform-samples}
Let $n,m,q \in \N, \sigma, r, \gamma \in \R$. Let $\S$ be a distribution over $\Z^n$ where all elements in the support have fixed norm $r$. 
Suppose there is no $T + \poly(n,m, \log(\secp), \log(q))$ time distinguisher between the distributions $\CLWE(m, D_1^n, \frac{1}{r} \cdot \S, \gamma, \beta)$ and $D_1^{n \times m} \times U(\T_1^m)$. Then, there is no $T$-time distinguisher between the distributions $\LWE(m, \T_q^n, \S, D_\sigma)$ and $U(\T_q^{n \times m} \times \T_q^m)$, with an additive advantage loss of $\negl(\secp)$, where
\begin{align*}
    \gamma &= r \cdot \sqrt{\ln n + \ln m + \omega(\log \secp)},
    \\\sigma &= \beta \cdot q.
\end{align*}
\end{lemma}

\begin{proof}
Suppose we have one sample $(\veca, b)$, from either $\CLWE(m, D_1^n, \frac{1}{r} \cdot \S, \gamma, \beta)$ or $D_1^{n \times m} \times U(\T_1^m)$. Now, consider the sample $(\veca \cdot \tau \cdot q \pmod{q}, b \cdot q \pmod{q})$, where $\tau = \sqrt{\ln(n) + \ln(m) + \omega(\log\secp)}$ and $\gamma = r \cdot \tau$. Let $\vecs \sim \frac{1}{r} \cdot S$ be the CLWE secret, and let $\veca' = \veca \cdot \tau \cdot q$, let $e' = e \cdot q$, and let $\vecs' = r \cdot \vecs \in \Z^n$. If $(\veca, b)$ is from the CLWE distribution, then we have (taking all components mod $q$)
\begin{align*}
(\veca \cdot \tau \cdot q, b \cdot q) &= (\veca', (\gamma \cdot \inner{\vecs, \veca} + e)  \cdot q)
\\&= (\veca', \gamma \cdot \inner{\vecs, q \cdot \veca} + e')
\\&= (\veca', \frac{\gamma}{r \cdot \tau} \cdot \inner{r \cdot \vecs, \tau \cdot q \cdot \veca} + e')
\\&= (\veca', \inner{r \cdot \vecs, \tau \cdot q \cdot \veca} + e')
\\&= (\veca', \inner{\vecs', \veca'} + e')
\\&= (\veca', \inner{\vecs', \veca' \pmod q} + e').
\end{align*}
Note that $\vecs' = \vecs \cdot r \sim \S$ and $e' = e \cdot q \sim D_{\beta \cdot q}$, so $\vecs'$ and $\vece'$ have the right distribution. Lastly, $\veca' = \tau \cdot q \cdot \veca \sim D_{\tau \cdot q}^n$, so by Lemma~\ref{lemma:gaussian-mod-lattice-looks-uniform}, $\veca' \pmod{q}$ is $\negl(\secp)/m$-close to $\T_q^n$ as long as $q \cdot \tau \geq \eta_{\negl(\secp)/m}(q \cdot \Z^n)$, which holds by Lemma~\ref{lemma:smoothing-parameter-estimate} by construction of $\tau$. Thus, taking the triangle inequality over all $m$ samples, the resulting distribution is $\negl(\secp)$-close to $\LWE(m, \T_q^n, \S, D_\sigma)$ where $\sigma = \beta \cdot q$.

Lastly, if $(\veca, b)$ is from the null distribution, then clearly $b \cdot q \sim \T_q$, and by the same argument as above, $\veca' \pmod{q}$ is $\negl(\secp)/m$-close to $\T_q^n$, which by the triangle inequality, implies the resulting distribution is $\negl(\secp)$-close to $U(\T_q^{n \times m} \times \T_q^m)$, as desired.
\end{proof}

\begin{lemma}[Reverse of Lemma~\ref{lemma:discrete-to-continuous-samples}]\label{lemma:continuous-to-discrete-samples}
Let $n, m, q \in \N$, $\sigma \in \R$. Let $\S$ be a distribution over $\Z^n$ where all elements in the support have fixed norm $r$, and suppose that
\[\sigma \geq 3r \sqrt{\ln n + \ln m + \omega(\log \secp)}. \]
Suppose there is no $T + \poly(m, n, \log(\secp), \log(q))$-time distinguisher between the distributions $\LWE(m, \T_q^n, \S, D_{\sigma})$ and $U(\T_q^{n \times m} \times \T_q^m)$. Then, there is no $T$-time distinguisher between the distributions $\LWE(m, \Z_q^n, \S, D_{\sigma'})$ and $U(\Z_q^{n \times m} \times \T_q^m)$ with an additive $\negl(\secp)$ advantage loss, where we set
\[ \sigma' = \sqrt{\sigma^2 + 9r^2 (\ln n + \ln m + \omega(\log \secp))} = O(\sigma). \]
\end{lemma}

\begin{proof}
Suppose we are given a sample $(\veca, b)$ from either $\LWE(m, \T_q^n, \S, D_{\sigma})$ or $U(\T_q^{n \times m} \times \T_q^m)$. Let $\veca' \sim D_{\Z^n - \veca, \tau}$, where $\tau = \sqrt{\ln(n) + \ln(m) + \omega(\log \secp)}$. Let $\veca'' = \veca + \veca' \pmod{q}$, and observe that $\veca'' = \veca + \veca' \pmod{q} \in \Z_q^n$, as $\veca'$ is supported on $\Z^n - \veca$. Now, consider the sample $(\veca'', b)$. Let $\vecs \sim \S$ be the LWE secret. If this is from the LWE distribution, we have
\begin{align*}
(\veca'', b) &= (\veca'', \inner{\vecs, \veca} + e)
\\&=(\veca'', \inner{\vecs, \veca'' - \veca'} + e)
\\&=(\veca'', \inner{\vecs, \veca''} - \inner{\vecs, \veca'} + e)
\\&=(\veca'', \inner{\vecs, \veca''} + e'),
\end{align*}
where we define $e' = e - \inner{\vecs, \veca'}$. First, let's analyze the distribution of $e'$. By applying Lemma~\ref{lemma:discrete-plus-continuous}, since $\vecs$ has norm $r$, we know that $e'$ is $\negl(\secp)/m$ close to $D_{\sigma'}$ where
\[ \sigma' = \sqrt{\sigma^2 + r^2 \tau^2} = \sqrt{\sigma^2 + r^2(\ln(n) + \ln(m) + \omega(\log \secp)},\]
as long as
\[ \eta_{\negl(\secp)/m}(\Z^n) \leq \frac{1}{\sqrt{1/\tau^2 + (r/\sigma)^2}}, \]
which holds if $\tau, \sigma/r \geq \sqrt{2} \cdot \eta_{\negl(\secp)/m}(\Z^n)$, which it does by Lemma~\ref{lemma:smoothing-parameter-estimate} and construction of $\tau$ and our condition on $\sigma$. Therefore, by the triangle inequality over all $m$ samples, the errors look $\negl(\secp)$-close to $D_{\sigma'}$.

Now, we consider the distribution of $\veca'' = \veca + \veca' \pmod{q} \in \Z_q^n$. Note that the lattice $\Z^n + \veca$ depends only on $\veca \pmod{1}$, so given $\Z^n + \veca$, the conditional distribution on $\veca$ is $U(\Z_q^n + (\veca \pmod{1}))$. Therefore, by a one-time pad argument, the distribution of $\veca + \veca'$ is exactly $U(\Z_q^n)$, even conditioned on $\veca'$. Therefore, $(\veca'', \inner{\vecs, \veca''} + e')$ looks $\negl(\secp)/m$-close to a sample from $\LWE(m, \Z_q^n, \S, D_{\sigma'})$, which by the triangle inequality over $m$ samples, makes the resulting distribution $\negl(\secp)$-close to $\LWE(m, \Z_q^n, \S, D_{\sigma'})$.

Lastly, suppose $(\veca, b)$ is from the null distribution. Then by (a simpler version of) the above argument, the distribution of $\veca''$ is given by $\veca'' \sim U(\Z_q^n)$, making $(\veca, b) \sim U(\Z_q^n \times \T_q)$, as desired.
\end{proof}

Now we are ready to prove Theorem~\ref{thm:clwe-to-lwe}.

\begin{proof}[Proof of Theorem~\ref{thm:clwe-to-lwe}]
Suppose there is no $T + \poly(n, m, \log(q), \log(\secp))$ time distinguisher between $\CLWE(m, D_1^n, \frac{1}{r} \cdot \S, \gamma, \beta)$ and $D_1^{n \times m} \times U(\T^m)$ with advantage at least $\epsilon - \negl(\secp)$. Then, by Lemma~\ref{lemma:gaussian-to-uniform-samples}, there is no distinguisher between $\LWE(m, \T_q^n, \S, D_{\sigma})$ and $U(\T_q^{n \times m} \times \T_q^m)$, where $\gamma = r \sqrt{\ln(n) + \ln(m) + \omega(\log \secp)}$ and $\sigma = \beta \cdot q$. Then, by Lemma~\ref{lemma:continuous-to-discrete-samples}, there is no $T$-time distinguisher between $\LWE(m, \Z_q^n, \S, D_{\sigma'})$ and $U(\Z_q^{n \times m} \times \T_q^m)$ with advantage $\epsilon$, where $\sigma' = O(\sigma)$, as long as $\sigma = \beta \cdot q \geq 3r \sqrt{\ln n + \ln m + \omega(\log \secp)}$.
\end{proof}